**Title:** The Basic B**** Effect: The Use of AI Agents Reduces the Distinctiveness and Diversity of People's Choices.

**Authors:**

Matz*, Sandra C.; Columbia Business School

Klugescheid, Kimberly; Columbia Business School

Horton, C. Blaine; Columbia Business School

Goethals, Sofie; University of Antwerp

*Corresponding author: sm4409@gsb.columbia.edu, +1 3479786326

**Author Contributions:** S.C.M developed the research idea. S.C.M and K.K. collected and analyzed the data. S.C.M wrote the manuscript with feedback from K.K., C.B.H and S.G.

**Competing Interest Statement:** The authors have no competing interests to disclose.

**The Basic B**** Effect: The Use of AI Agents Reduces the Distinctiveness and Diversity of People's Choices.**

## Abstract

Large language models (LLMs) increasingly act on people's behalf: they write emails, buy groceries, and book restaurants. While the outsourcing of human decision-making to AI can be convenient, it raises a fundamental question: how does delegating identity-defining choices to AI shape who people become? Across a large field study and a controlled experiment, we study the impact of agentic LLMs on two identity-relevant outcomes: interpersonal distinctiveness—how unique a person's choices are relative to others—and intrapersonal diversity—the breadth of a single person's choices over time. Study 1 uses 110,000 real choices drawn from social media behavior of 1,000 U.S. users to compare generic and personalized agents to a human baseline. Both agents shift people's choices toward more popular options, reducing the distinctiveness of their preferences. While the use of personalized agents tempers this homogenization (compared to generic agents), it also more strongly compresses the diversity of people's preference portfolios by narrowing their exploration across topics and psychological affinities. Study 2 replicates these patterns in an online experiment which mimics common real-world scenarios (e.g., choosing movies) and allows us to directly compare the AI agent's choices to those made by 348 participants (12,097 human choices). The findings also suggest that the flattening effects of AI agents are amplified when choices are made sequentially (vs. batch), and when agents rely on domain-specific user information for personalization. Understanding how AI agents compress human experience (and the trade-offs involved) is critical for designing systems that augment human agency and safeguard diversity in thought, taste, and expression.

## Significance Statement

Agentic AI promises frictionless living, but subtle shifts in what gets chosen by people's agents can reshape who they become. Across a large field study and a controlled experiment, we show that LLM-based agents nudge behavior toward more normative options and narrow the range of what individuals explore. Our findings suggest that agentic AI, for all its convenience and efficiency, threatens to flatten the human experience and involves trade-offs between interpersonal distinctiveness and intrapersonal diversity.

*"My automated advisers saved me time and alleviated the burden of constantly making choices, but they seemed to have an agenda: Turn me into a Basic B."*

Kashmir Hill, New York Times Oped[1]

## Introduction

Large language models (LLMs) have rapidly become a cornerstone of everyday decision-making, influencing what people watch and buy, how they work and communicate, and even how they vote[2]. Increasingly, LLM-based systems are designed not just to advise people but to act on their behalf (i.e., agentic AI [3,4]). Instead of recommending a restaurant, an AI agent books it. Instead of suggesting a new outfit, an agent buys it. And instead of floating the idea of meeting a childhood friend for drinks, an agent automatically schedules it. This transition, from being passive tools to active agents that shape behavior through automated decisions, raises an important question: What —if anything—do we stand to lose when decisions that define who we are as individuals are delegated to AI?[5]

A growing body of work highlights the potential of LLMs to support more efficient, informed, and personalized decision-making across a wide variety of domains[2,6–8]. Yet little is known about the broader psychological and behavioral consequences of delegating decisions to such systems. In this paper, we investigate how the use of LLMs as personal agents influences two key aspects of human choice: *interpersonal distinctiveness* (how unique and distinct a person's behaviors and preferences are compared to those of the overall population) and *intrapersonal diversity* (how broad and diversified a single person's portfolio of behaviors and preferences are across time and space.

We hypothesize that the growing reliance on LLM-based agents will lead to a flattening of the human experience by dampening both the interpersonal distinctiveness and intrapersonal diversity of people's choices (see Fig. 1). LLMs are trained to detect statistical regularities in large-scale human data and produce outputs that are probabilistic[9]. As a consequence, when making decisions, they tend to err on the side of exploitation, selecting the most likely or statistically frequent option rather than introducing more novel or diverse alternatives. For example, one might reasonably expect an LLM to reply with a box-office sensation like *Titanic* or a culturally prominent film like

*The Notebook*, rather than a less well-known film such as *In Your Eyes*, when asked to complete the sentence, "A romantic drama that everybody should watch is...". We argue this tendency of LLM-based agents to favor popular, widely accepted choices will, when relied upon for decision-making, lead to more uniform preferences and behaviors in two ways: *across* different individuals and *within* the same individual over time.

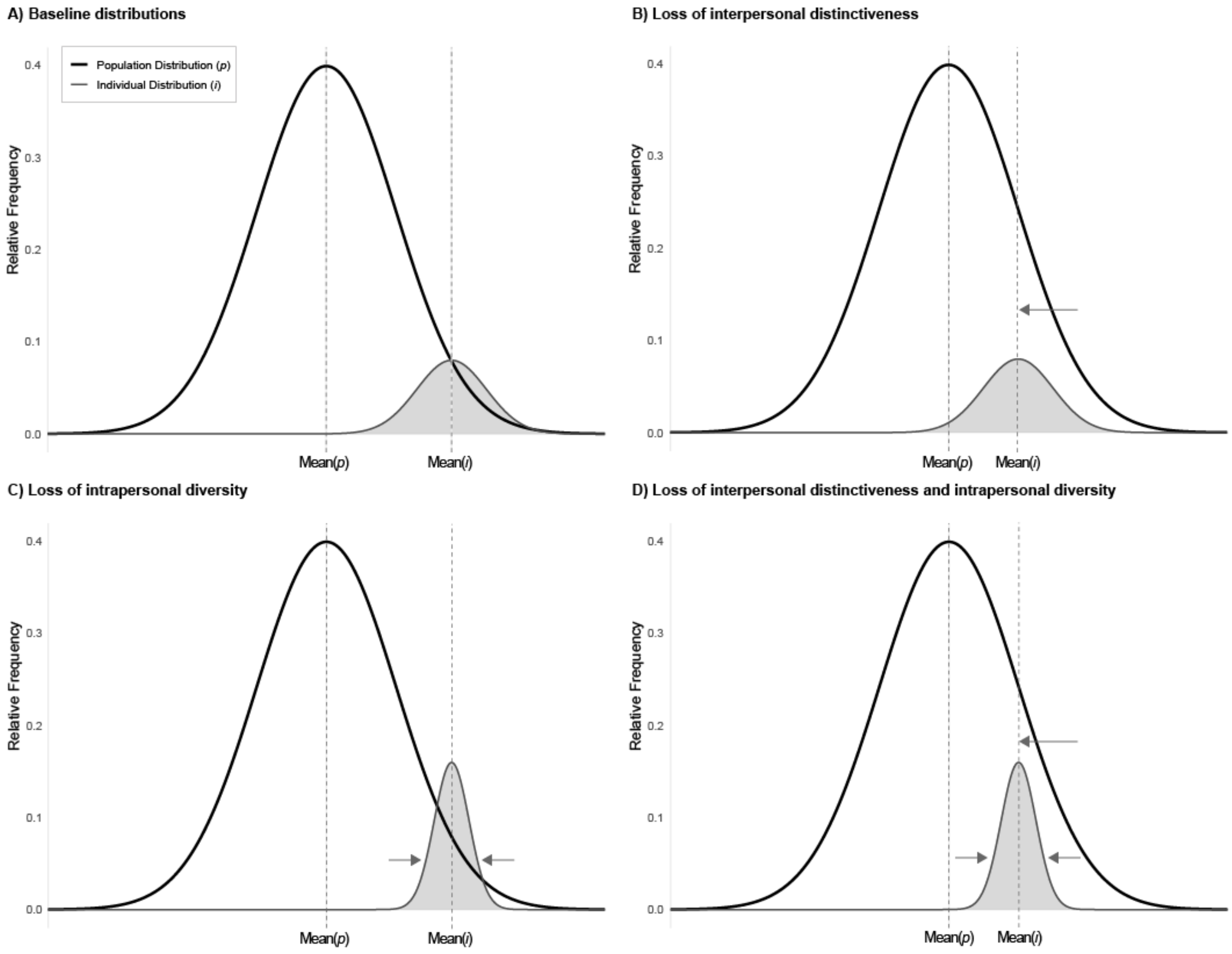

**Figure 1.** A) Two hypothetical preference distributions – one for the overall population $p$ and one for a particular individual $i$ within that population – in a baseline state. B) Loss of personal distinctiveness: The individual distribution is pulled towards the population mean. C) Loss of intrapersonal diversity: The variance of the individual distribution is shrinking. D) Loss of both interpersonal distinctiveness and intrapersonal diversity: the individual distribution is pulled towards the population mean and its variance is shrinking.

Emerging research supports this concern at the collective level. For instance, LLM-assisted creative work tends to become homogenized across users[10–14]. Relatedly, LLMs have been shown to significantly reduce the diversity and breadth of historical figures identified as most influential across different domains (e.g., sports, science, politics[15,16]). However, these findings primarily reflect between-person convergence—a reduction in variability across different users in tasks like content generation or factual queries—rather than a loss of variation within individuals over time in the context of preference-based decisions. That is, prior work has largely examined what happens to the collective when people collaborate with LLMs to create or answer, not what happens to the individual when they ask an LLM to choose for them.

This question is becoming more important as people increasingly rely on personalized agents which are explicitly intended to accurately reflect and reinforce (rather than shift or narrow) individual preferences (e.g., "digital twins"[17] designed to simulate a particular individual's preferences, personality, and choices). Recent work suggests that LLMs can infer and reproduce many aspects of an individual's existing preferences when provided with sufficiently rich behavioral histories[18,19]. At first glance, such personalization might appear to protect against the homogenizing effects of generic AI recommendations. We propose, however, that personalization alone may not preserve the breadth of an individual's decision-making. Instead, personal agents intended to preserve an individual's distinctiveness may actually lead to an even greater reduction in choice diversity by repeatedly optimizing for the most likely option within that individual's typical set of choices.

We test these predictions in two complementary studies. Study 1 is a large field study that uses real choices drawn from social media behavior of 1,000 Facebook users to compare the distinctiveness and diversity of people's choices in two AI agent conditions (generic and personalized) to a human baseline. Study 2 is a controlled online experiment that mimics multiple common real-world scenarios (e.g., selecting movies to watch or books to read), allowing us to directly contrast the distinctiveness and diversity of AI agents' choices to those made by participants under different conditions of personalization (e.g., random or personalized sorting of movies). Together, these studies examine whether delegating decisions to AI agents reduces both interpersonal distinctiveness and intrapersonal diversity, and whether personalization mitigates or amplifies those effects. The data, code and materials are available through our OSF repository.

## Results

**Study 1.** We examine over 110,000 real-world choices made by 1,000 individuals to test whether generic and personalized AI agents lower individuals' choice distinctiveness and diversity, when compared to their natural baseline. Our data comes from the myPersonality project, a Facebook application that allowed users to take psychometrics tests (e.g., personality) and share their Facebook profiles for research purposes[20].

For each user, we prompted GPT-4o to choose between two Facebook pages that—unknown to the LLM—were randomly drawn from a user's actual Facebook profile (i.e., both pages were followed by the sampled user). Facebook pages are an integral part of the platform, allowing users to express enduring interests by subscribing to public figures, organizations, brands, and other entities. Here, we treat the user's decisions to follow Pages as a revealed-preference proxy for everyday tastes. Following the official Facebook page of Lady Gaga or the *New York Times*, for example, is a likely manifestation of a preference that extends beyond the platform and into the "offline" world (e.g., listening to Lady Gaga or reading the *New York Times*).

ChatGPT was either prompted in the form of a generic agent (i.e., "*Which of the following two Facebook pages would you recommend I follow next?*") or a personalized agent. For each personalized agent, we added information about a user's age, gender, and a selection of 10 random Facebook pages that user followed (separate from the actual choice sets), as well as a 20-line summary of their preferences. The 20-line summary of the participant was automatically created based on the user's status updates (average number of status updates per user: 347 ± 254 *SD*). Specifically, we prompted GPT-4o to generate a summary profile of the user with the following instructions: *"You are a behavioral scientist and expert profiler. Please describe the user who wrote the following Facebook status updates in no more than 20 lines."* **Fig. 2** shows an example prompt that is based on partial data of an actual participant in the dataset.

**Personalized AI**

*"Below is a description of a Facebook user:*

*The user's gender is: female*

*The user's age is: 27*

*The user likes these 10 pages on Facebook:*

- *The Twilight Saga*
- *Taylor Swift*
- *Road Trips*
- *The Simpsons*
- *Breastfeeding*
- *...*
- *Gummy Worms*

*The user has the following personal characteristics:*

- ***Family-Centric:*** *The user places considerable emphasis on family life, often mentioning their spouse and children. They frequently engage in family activities and express love and appreciation for their loved ones.*
- ***Social and Community-Oriented:*** *The user enjoys socializing, attending events like volleyball games and birthday parties, and spending time with friends and extended family. They often mention nights out with friends, and express a fondness for such interactions.*
- ***Emotionally Expressive:*** *The user is open about their emotions, sharing both their joys and frustrations. They express love towards their family and friends but also mention feeling bored, dealing with drama, and facing work-related stress.*
- *...*
- ***Active Lifestyle:*** *Engages in various activities, including volleyball, attending festivals, and family outings to places like lakes and parades. They also partake in domestic activities like house cleaning and shopping.*

*Based on the description of this user, which of the following two Facebook pages would they prefer to follow next:*

*Page 1: Beyoncé*

*Page 2: Grilling Out*

*Think carefully and consider what aligns most with their preferences."*

**Fig 2**. Prompt used to instruct the personalized version of ChatGPT to choose between two of the participant's Facebook pages. User information was dynamically piped into each prompt. Note, the example above only shows partial information of an example participant.

Each AI agent (generic and personalized) made a total of 50 binary choices. To establish a baseline of observed human preferences, we randomly selected one of the two pages for each of the 50 pairs. Although the selection from each pair was random, both options reflected preferences previously expressed by that user. Thus, this baseline is not intended to reconstruct the user's original decision process; rather, it provides a reference distribution grounded in the user's revealed preferences, absent any agent intervention, against which the consequences of AI selection can be evaluated. This procedure resulted in three matched choice sets: one containing 50 pages selected by the generic agent ("Generic AI"), one containing 50 pages selected by the personalized agent ("Personalized AI"), and one containing 50 pages sampled from the user's observed preferences ("Observed Human Preferences"). See **Fig. S1** in Appendix A of the SI for an illustration of the analytical procedure.

We conducted paired *t*-tests to compare the distinctiveness of choices—measured as the average inverse popularity of pages chosen—across all three choice sets (see **Fig. 4**). As predicted, the choice sets curated by the AI agents showed lower distinctiveness scores (generic agent: $t(999) = -11.02$, $p < .001$, $d = 0.21$; personalized agent: $t(999)$ = =-4.08, $p < .001$, $d = 0.08$), indicating that the use of AI agents resulted in a selection of less unique and more widely shared preferences relative to the observed preference baseline. Notably, the effect was over 2.5-times larger for the generic AI agent ($t(999) = 7.10$, $p < .001$, $d = 0.13$), highlighting the potential of personalized agents to buffer against the homogenization of preferences in the population.

However, the tendency of personal agents to preserve higher levels of choice distinctiveness appears to come at the cost of choice diversity. We measured the choice diversity for each user in two ways: (1) *topical diversity*, defined as the extent to which the chosen Facebook pages were evenly spread across different categories like sports, music or TV, and (ii) *psychological diversity*, defined as the extent to which the chosen Facebook pages showed diversity in the psychological profiles of their followers (the two metrics are largely uncorrelated and therefore distinct; see Methods for more details). The use of a personal agent led to a consistent and notable drop in choice diversity, both when measured as topical diversity ($t(999) = -16.71$, $p < .001$, $d = -0.31$) and psychological diversity ($t(999) = -5.69$, $p < .001$, $d = -0.09$). The generic agent, in contrast, reduced topical diversity to a significant but smaller degree than the personal agent ($t(999) = -10.80$, $p <$

.001, $d$ = -0.21), and, in fact, increased psychological interest diversity compared to the human baseline ($t(999) = 7.62$, $p < .001$, $d = 0.12$).

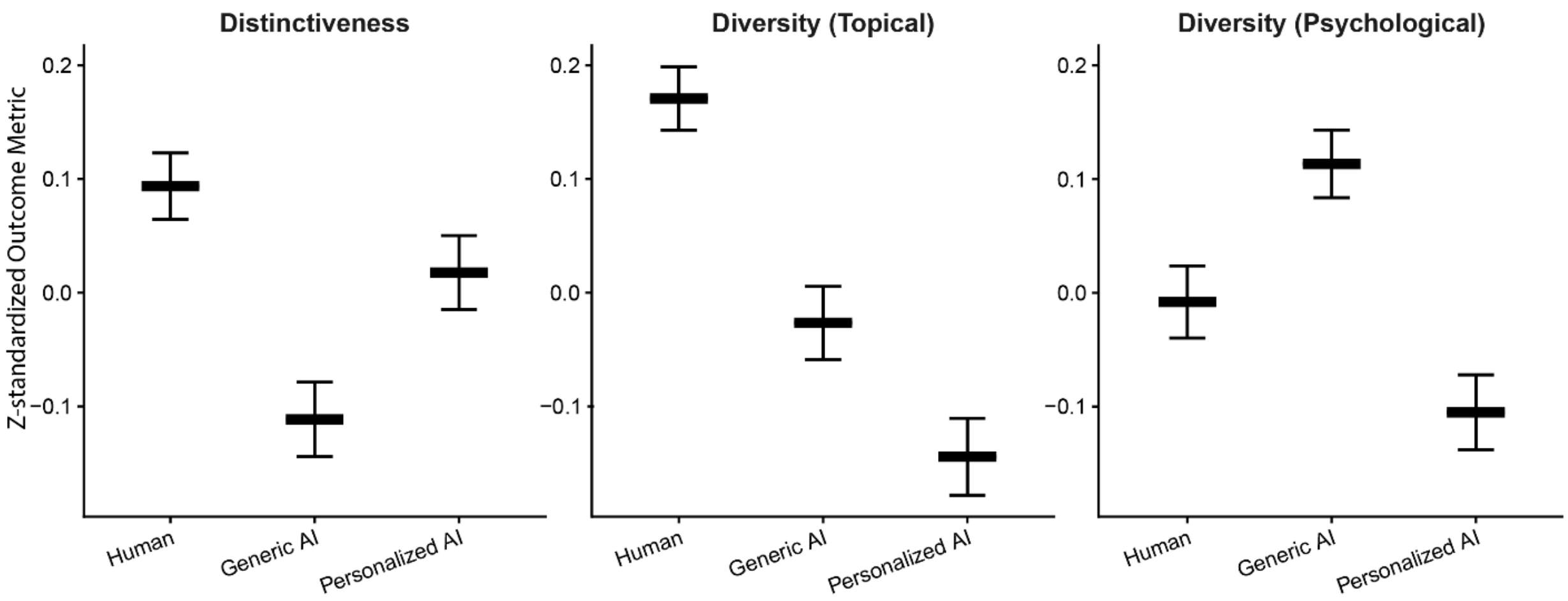

**Fig. 4.** Average scores (and standard errors) for choice distinctiveness (A) and choice diversity outcomes (B and C). All scores were $z$-standardized within each metric before being averaged.

**Study 2.** We replicate and extend the findings of Study 1 by examining over 10,000 choices made by 348 participants in the context of a web application that mimicked popular recommendation platforms across three choice domains: movies, books and podcasts (see **Fig. 5** for the example of the movie platform).

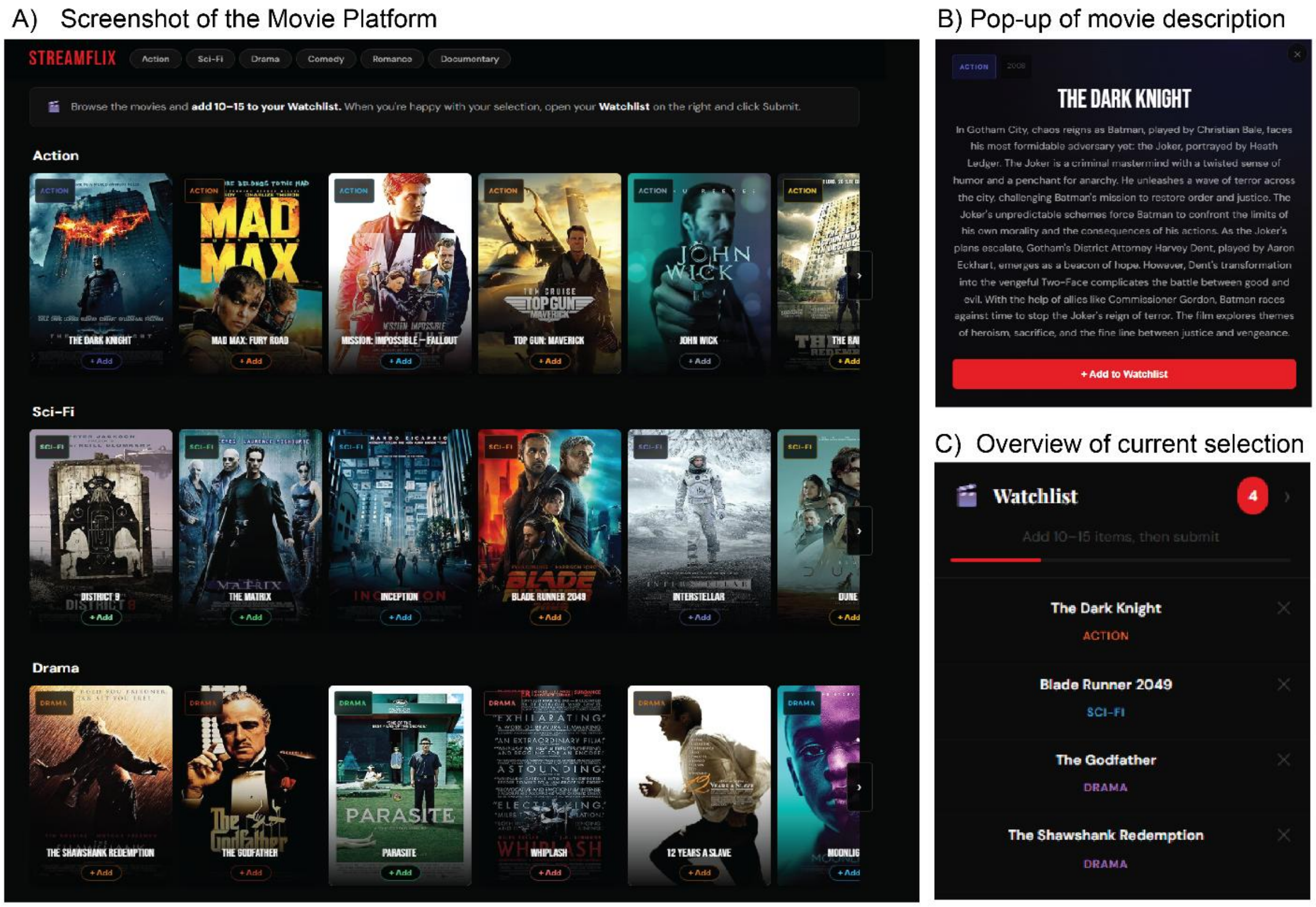

**Fig. 5.** Screenshots of the movie platform.

We deployed a within-person experimental design in which participants browsed a catalog of either 150 movies, books, or podcasts and added between 10 and 15 items to a personal watchlist, shopping basket, or playlist, depending on the domain. Participants were free to browse and revise their selections before submitting their final collection. Items were displayed under 6-7 distinct categories (e.g., Action, Sci-Fi, or Drama for movies). Each participant was randomly assigned to one of three experimental conditions that mimicked various degrees of personalization in how the available options were organized: 1) a random order condition, in which all categories and items within categories were displayed in random order, 2) a generic recommendation condition, in which categories and items within categories were sorted by a generic instance of ChatGPT, and

3) a personalized recommendation condition, in which categories and items within categories were sorted by a personalized instance of ChatGPT. Importantly, this design did not limit the number of choices available to participants (i.e., each individual could see all 150 options). Instead, conditions simply influenced how salient each item was based on the different personalization strategies.

Before comparing human choices to those made by AI agents, we tested whether the three conditions significantly differed in the distinctiveness of their choices (measured as inverse popularity) and diversity of choices (measured as entropy for topical diversity and the average cosine distance of item embeddings for psychological diversity). As **Tables S1-3** in Appendix B in the SI show, there were no significant differences across conditions in any of the outcome metrics (except for a small difference between the random and the personalized conditions on distinctiveness). In other words, the order in which movies/books/podcasts were presented to participants did not significantly impact how unique and varied their ultimate choices were. Thus, we considered all three conditions as one "Human Choice" condition, which allowed us to utilize the full sample for each choice domain.

The distinctiveness and diversity of the Human Choice condition served as the comparison baseline for the choices made autonomously by two agents (which received access to the originally index list of options and were prompted using a strategy similar to the one employed in Study 1, see Methods for more details). A generic agent was prompted to select the *n* best movies for a participant without any additional information about the participant, where *n* corresponds to the actual number of items selected by the participants. A personalized agent, in contrast, received access to a 20-line summary that was generated based on the participant's ChatGPT history.

**Fig. 6** displays the average choice distinctiveness and diversity scores (*z*-standardized within choice domains) for the Human Choice condition as well as the two AI agent conditions. To facilitate interpretation, the X-axis in each graph represents the level of distinctiveness and diversity when the movies/books/podcasts for each participant were selected at random.

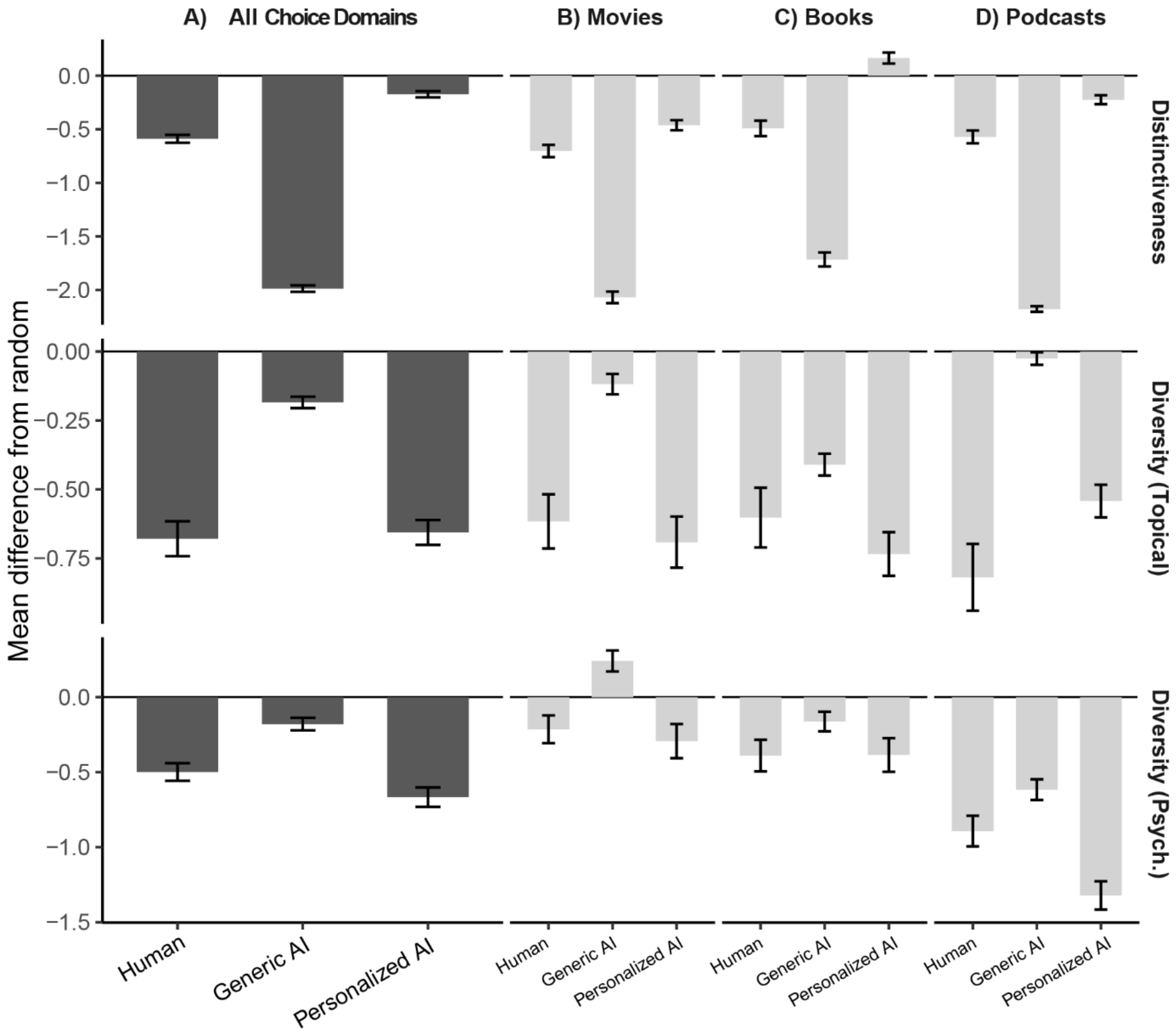

**Fig 6.** Average scores (and standard errors) for choice distinctiveness (row 1) and choice diversity outcomes (rows 2-3), collapsed across all choice domains (column A) and separately for each decision domain (columns B-D). All scores were *z*-standardized within outcome and domain before being averaged. The y-axis intercept is set to the outcome for the random condition, allowing for a direct comparison across all metrics and contexts.

We ran a series of multilevel models predicting the distinctiveness and two diversity metrics from the choice condition (Random Selection, Human Choice, Generic Agent and Personalized Agent) with observations cross-nested in both the choice domain and participant ID (see **Tables S4-6** in Appendix C in the SI for full model outputs). As **Fig. 6** illustrates, the distinctiveness and diversity of the choices made by participants and their two agents were significantly lower than those observed for the random selection of options – represented by the X-axis in **Fig. 6** (see Column A for results collapsed across choice domains and Columns B-D for individual findings). As in Study 1, the generic agent led to a starker reduction in distinctiveness ($\beta$ = -2.05, SE = 0.03, $t$ = -72.71, $p$ < .001, $d$ = 3.18) relative to the personalized agent ($\beta$ = -0.15 SE = 0.03, $t$ = -5.49, $p$ < .001, $d$ = 0.24), whereas the personalized agent resulted in a bigger drop in diversity (topical: $\beta$ = -0.37, SE = 0.04, $t$ = -9.88, $p$ < .001, $d$ = 0.43; psychological: $\beta$ = -0.62, SE = 0.04, $t$ = -15.74, $p$ < .001, $d$ = 0.69) relative to the generic agent (topical: $\beta$ = -0.10, SE = 0.04, $t$ = -2.74, $p$ < .001, $d$ = 0.12; psychological: $\beta$ = -0.20, SE = 0.04 , $t$ = -5.00, $p$ < .001, $d$ = 0.22).

Unlike Study 1, where a behavioral baseline was constructed by sampling from options that participants had already chosen (i.e., their revealed preference set), the baseline in Study 2 randomly samples from all available options in the environment. It therefore serves as a theoretical upper bound on the distinctiveness and diversity attainable if selections were unconstrained by either human or AI preferences. Perhaps unsurprisingly, both human participants and AI agents produced choices that were less distinctive and less diverse than this benchmark. Consequently, the comparisons between the Human Choice condition and the two AI agent conditions provide the more behaviorally meaningful test of whether delegating decisions to AI further reduces distinctiveness and diversity beyond ordinary human choice

Replicating our finding in Study 1, the choices curated by the generic AI agent showed significantly lower distinctiveness scores than those made by humans ($\beta$ = -1.42, SE = 0.03, $z$ = -50.33, $p$ < .001, $d$ = -2.20), indicating that the use of generic AI agents resulted in a selection of less unique and more widely shared preferences. In contrast, the use of personalized AI agents led to an increase in distinctiveness compared to participants' actual selection ($\beta$ = 0.48, SE = 0.03, $z$ = 16.89, $p$ < .001, $d$ = 0.74), suggesting that personalized agents might not only buffer against (see results of Study 1) but in some instances even support the exploration of less popular options.

However, mirroring the results of Study 1, the benefit of personal agents in the context of choice distinctiveness came at the cost of choice diversity. While the generic agents, in fact, significantly increased choice diversity compared to the human baseline (topical diversity: $\beta$ = 0.25, SE = 0.04, $z$ = 6.69, $p$ < .001, $d$ = 0.29; psychological diversity: $\beta$ = 0.26, SE = 0.04, $z$ = 6.53, $p$ < .001, $d$ = 0.29), personal agents significantly decreased psychological diversity ($\beta$ = -0.17, SE = 0.04, $z$ = 4.22, $p$ < .001, $d$ = -0.18) while being on par with human choices for topical diversity ($\beta$ = 0.02,

SE = 0.04, $z$ = 0.44, $p$ = 1.00, $d$ = -0.02). In other words, personalization helped people remain more distinct from one another, but at the cost of narrowing the breadth of their own choices.

Although our primary analyses demonstrate that delegating decisions to AI changes the distinctiveness and diversity of people's choices, they leave open an important practical question: Under what conditions—and, in particular, which design choices—are the flattening effects of AI agents amplified or mitigated?

To explore this question, we manipulated two design features of the recommendation process. First, we compared agents that selected all $n$ items simultaneously (batch recommendations, akin to the original architecture) with agents that selected one item at a time (sequential recommendations). Second, we varied the information available to personalized agents, contrasting domain-general and domain-specific sources. Specifically, we compared the original architecture where agents received a domain-general summary of the participant derived from their ChatGPT conversations, two a condition where the AI agents received the participant's actual choices within the focal domain (e.g., the movies they had selected), as well as a hybrid condition where the AI agents received access to both. To assure equivalence across conditions, the items selected by the participants were removed from the list of items the agents could chose from.

As **Fig 7.** shows, the loss of distinctiveness and diversity was amplified when recommendations were made sequentially (as opposed to batch-wise) and when personalized agents relied on domain-specific (rather than domain-general) information (see **Fig. S2** in Appendix D for a breakdown across domains, **Tables S4-S6** in Appendix C in the SI for the results of multilevel models and **Tables S7-S9** in Appendix E for pairwise contrasts). These findings suggest that the flattening effects of AI agents on distinctiveness and diversity are sensitive to the design choices that determine how the agents make their choices.

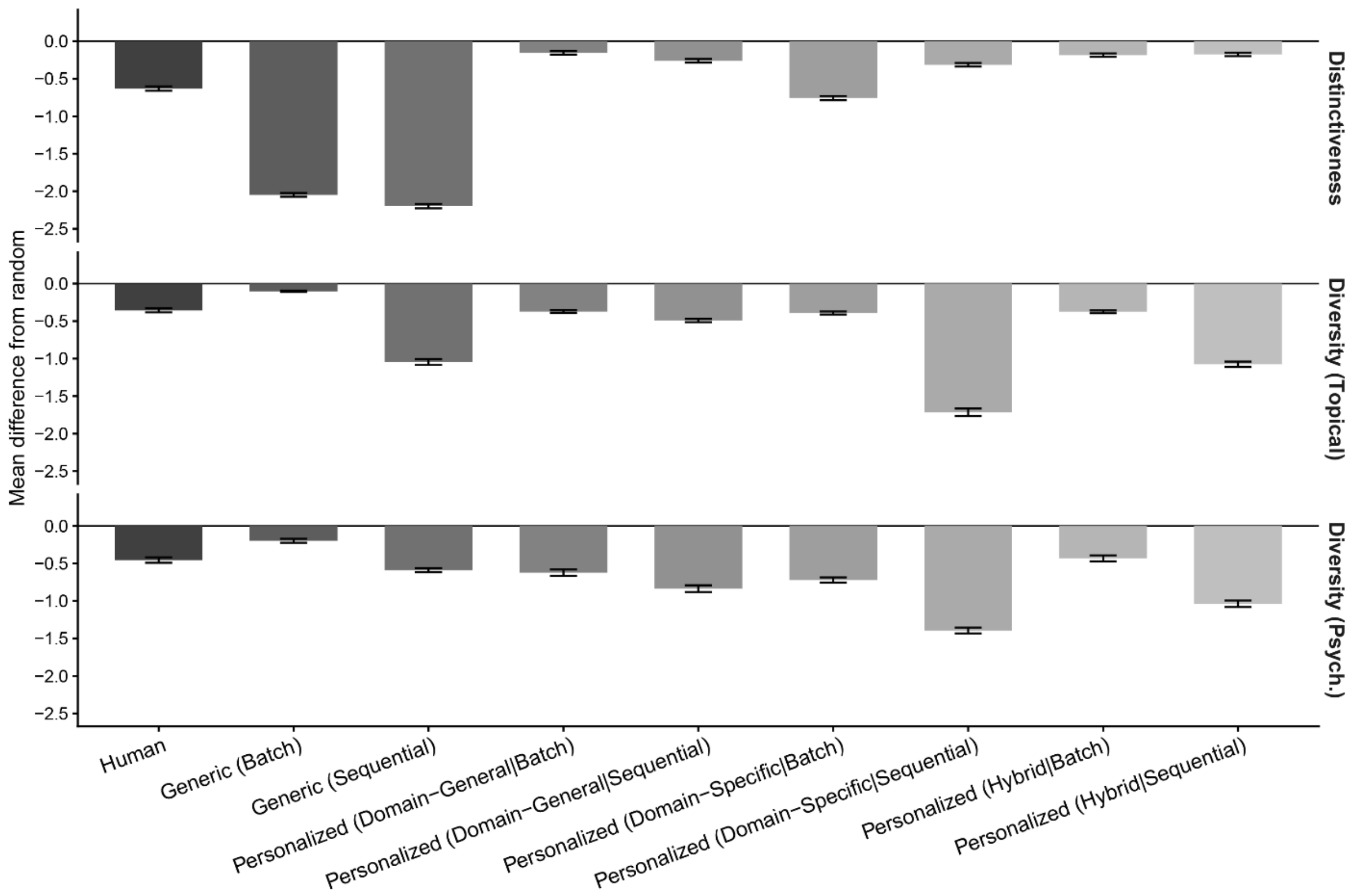

**Fig. 7.** Average scores (and standard errors) for choice distinctiveness (row 1) and choice diversity outcomes (rows 2-3), collapsed across all choice domains. All scores were *z*-standardized within outcome and domain before being averaged.

The preceding analyses examined whether different prompting strategies shift the overall level of distinctiveness and diversity in AI-generated choices. In a final set of exploratory analyses, we investigate a complementary question: whether personalized AI agents preserve individual differences in these properties of choice. Specifically, we examined whether participants who themselves made relatively more distinctive or diverse choices also received relatively more distinctive or diverse choice sets from their personalized agents. These correlations capture rank-order fidelity rather than absolute fidelity, testing whether AI agents may be able to preserve people's relative preferences of distinctiveness and diversity, even when they lower their absolute levels. As **Fig. 7** suggests, most personalized agents showed little evidence of preserving individual differences in how distinctive or diverse participants' choices were. Participants who themselves made relatively unusual or varied choices were generally no more likely to receive unusual or varied choice sets from their agents. The only exception was personalization based on domain-specific behavior—particularly when agents made choices in batches—which showed substantially greater rank-order correspondence with human choice patterns. More broadly, these findings suggest that the flattening of the human experience by AI agents may not only involve compression—shifting people toward less distinctive or diverse choices—but also decoupling, such that the choices made on a person's behalf become less reflective of their own characteristic behavioral patterns.

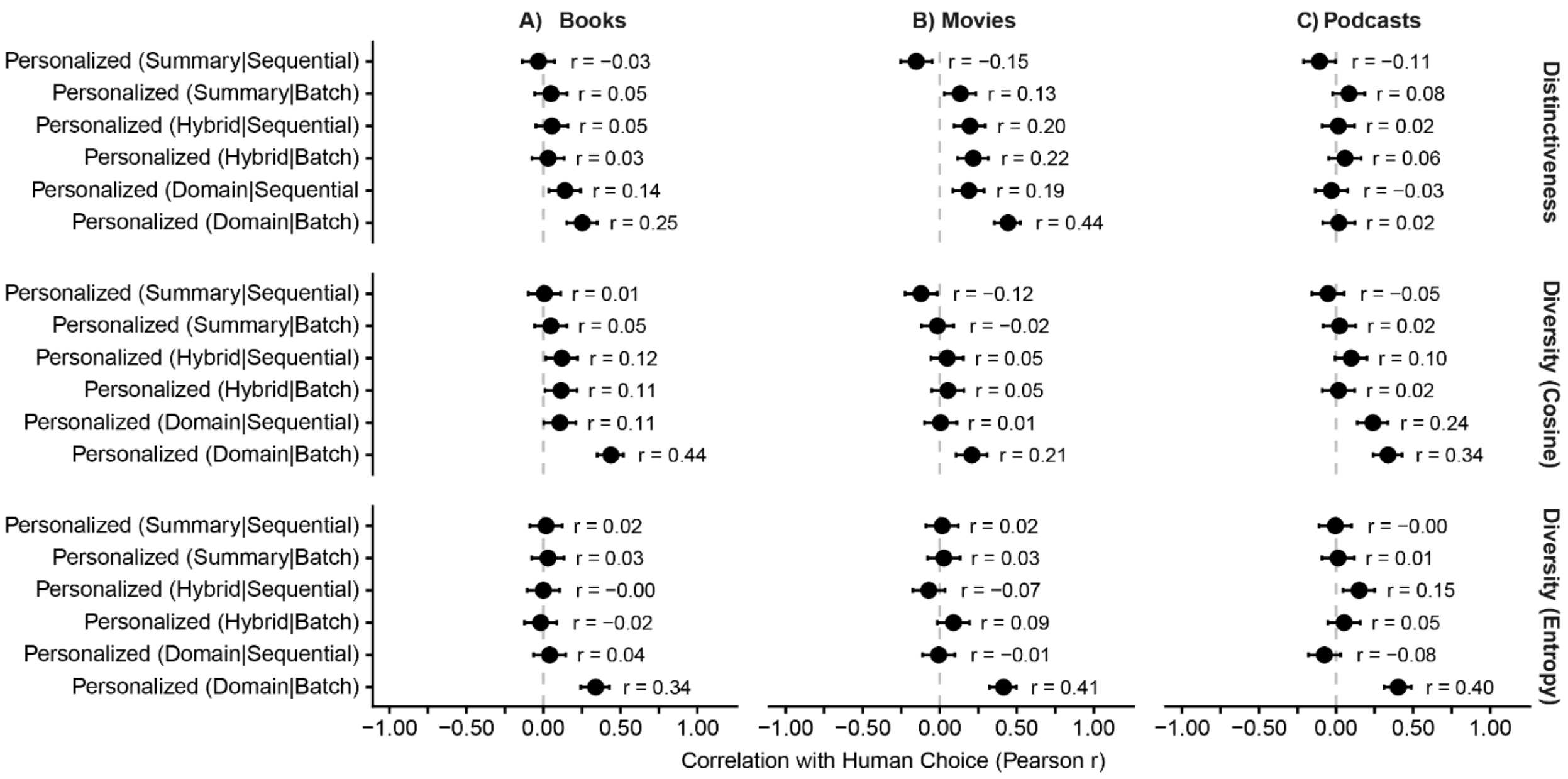

**Fig. 8.** Zero-order correlations between the choice distinctiveness (row 1) and diversity (rows 2-3) levels of the choices made by humans and their personalized AI agents. Higher correlations indicate that an agent more faithfully preserves individual differences in choice distinctiveness (top row), psychological diversity (middle row), and topical diversity (bottom row).

Overall, the findings from Study 2 suggest that—when compared to people's choices under real-world conditions—AI agents have the potential to both reduce and increase the distinctiveness and diversity of choices, but they do not do so simultaneously. While generic AI agents tend to flatten the distinctiveness of choices but amplify their diversity, personal AI agents flatten diversity but amplify distinctiveness. The findings also suggest that both losses become exaggerated when the AI makes choices in a sequential order (rather than batch) and when the information provided to the personalized AI is domain-specific (rather than domain-general).

## Discussion

Our findings reveal an important consequence of AI-assisted decision-making: while LLMs may offer convenience and consistency, delegating choices to them can systematically reshape both how distinct people are from one another and how varied their own choices are over time[5,21]. Critically, these two forms of heterogeneity do not necessarily move together. Generic LLMs show a strong gravitational pull toward choices that are widely shared in the population, thereby reducing interpersonal distinctiveness. Personalized LLMs, on the other hand, show a stronger tendency to concentrate choices within an individual's existing preference profile, thereby reducing intrapersonal diversity. This trade-off is especially clear in Study 2, where AI-generated choices could be compared directly with participants' new choices: generic agents produced choices that were less distinctive but more diverse, whereas personalized agents produced choices that were more distinctive but less diverse. Thus, personalization does not simply determine how strongly AI agents flatten human choice; rather, it influences which dimension of human heterogeneity is more likely to be preserved (an in some cases even elevated) or diminished.

These findings extend emerging evidence that LLM use can increase convergence across users by highlighting a potentially consequential feature of agentic AI: delegation. Prior research has largely examined settings in which LLMs generate ideas, answers, or recommendations that remain subject to human evaluation and selection[10–13,22]. When AI agents instead choose on a person's behalf, model-generated outputs can translate more directly into the experiences, products, and media that shape that person's behavioral portfolio. Thus, the consequences of LLM convergence may depend not only on what models generate, but also on the role those generations play in determining human behavior. An important direction for future research is therefore to examine whether the effects observed here vary along a continuum of delegation, from AI recommendation, to collaborative human–AI choice, to autonomous agentic selection.

Exploratory analyses further suggest that these effects depend not only on whether an agent is personalized, but also on how personalization is implemented. Losses in distinctiveness and diversity were amplified when agents made choices sequentially rather than in batches and when personalization relied on domain-specific rather than domain-general information. At the same time, domain-specific agents were generally better at preserving individual differences in participants' underlying levels of distinctiveness and diversity. This highlights an important tension between fidelity and breadth: agents that more closely reflect an individual's existing pattern of preferences may not necessarily preserve the full range of choices that individual would otherwise make. Our findings should be interpreted in light of several limitations. First, our studies primarily examine settings in which choices are made either by people or by AI agents. In some settings, human–AI decision-making may be more interactive, with users requesting recommendations, refining them through conversation, and ultimately retaining control over the final choice. The three human baseline conditions we deployed in the design of Study 2 mimic this interaction in the context of traditional recommender systems. However, in most cases that involve AI-based suggestions, a user might consult dynamically with their AI agent, ask for a few suggested options and go back and forth to iteratively improve the recommendations, but then ultimately decide by themselves which option to choose. While recent work has suggested that people oftentimes follow the recommendations made by AI[23], potentially blurring the distinction between AI recommendation and AI delegation, future research should examine whether such collaborative decision-making amplifies or mitigates the effects on distinctiveness and diversity observed here.

Second, our AI agents were exclusively deployed using LLMs developed by OpenAI (namely OpenAI's ChatGPT4.o in Study 1 and GPT-5-mini in Study 2), as GPT models still preserve the largest market share among individual users. Although LLMs often show remarkably similar properties in their behavior[24,25] future research should investigate how model behavior changes over time as LLMs become more complex and powerful (i.e., how are distinctiveness and diversity impact across different versions of ChatGPT), as well as compare OpenAI's behavior to those of other LLMS (e.g. Claude or Gemini).

More broadly, our findings challenge the assumption that faithfully representing a person means consistently selecting what is most characteristic of their past behavior. Human preferences are reflected not only in what people typically choose, but also in the breadth and variability of their choices across occasions. Personalization that accurately captures a person's central tendencies may therefore still fail to preserve the variability that characterizes their behavior. In other words, preserving human complexity may require agents to represent not only what a person is most likely to choose, but also how—and how much—that person is like to vary and explore. An important extension for future research will be to examine whether this challenge differs across individuals and demographic or cultural groups, such that personalization preserves some forms of human variation more faithfully than others.

In sum, our findings highlight an important risk: the gradual flattening of what makes individuals unique and dynamic when outsourcing identity-relevant choices to AI agents. We uncover an important trade-off between generic and personalized agents when it comes to the loss of either interpersonal distinctiveness or intrapersonal diversity. Understanding both the general risks of agentic AI as well as the specific trade-offs associated with different types of agents is critical for designing systems that augment rather than constrain human agency. Embedding distinctiveness and diversity as design objectives[26,27] (e.g., user-controlled exploration dials or diversity-aware optimization) and allowing users to specify which of those they want to optimize for in a given query, for example, could offer a practical path to preserving diversity in thought, taste, and expression in an increasingly algorithmically curated world.

## Methods

### Study 1.

**Data.** Our analysis is based on data from the myPersonality project, a Facebook application active between 2007 and 2012, which allowed users to take psychometrics tests (e.g. personality) and get immediate feedback on their responses. Users voluntarily granted the application access to their Facebook profiles for research purposes.

The myPersonality dataset consisted of 134,302 users for whom both Facebook pages and status updates were available. We arrived at our final analysis sample of 1,000 users based on the following steps: (1) We retained all U.S. users for whom information about age, gender and the Big Five personality traits was available. (2) We retained users who followed at least 110 Facebook pages (a requirement for the subsequent analyses, with 10 pages used to create the personal agent, and 100 pages used for the decision task). During this step, we only considered Facebook pages that were followed by at least 50 people and that were associated with a category (e.g. travel/leisure, sports, music), as reported by Facebook, which appeared at least 50 times. (3) We retained users who had posted at least 100 status updates. This selection procedure resulted in 4880 potential participants who all met our selection criteria. We then deliberately stratified our sample by age—rather than relying on a simple random draw—to ensure a more balanced representation of older users. Specifically, we sampled all available 182 users aged 41 and older, all available 227 users aged between 31 and 40, a set of 300 random users aged 25-30 (due to a coding error no 25-year-olds were sampled), and a set of 291 random users aged 18-24.

For each of the 1,000 participants, we randomly extracted 100 of their Facebook pages and grouped them into 50 binary choice sets (i.e., a comparison of two of their Facebook pages). This resulted in a total of 50,000 choices made by each AI agent. After excluding invalid responses where at least one of the AI agents did not accurately identify one of the two pages by name, 42,003 observations were retained for the final analysis (~42 per participant). It is noteworthy that the majority of cases in which the AI agent failed to choose a single page occurred in the generic AI condition.

**LLM Prompting.** Prompts are displayed in the results section. We accessed GPT-4o through OpenAI's API, keeping all parameters set to their default settings. This includes the parameter of temperature, which allows users to control the degree of randomness in the LLM's output distribution. We retained the default settings to estimate the impact of agents under typical use, where users are unlikely to modify these settings.

**Measures.** ***Choice Distinctiveness.*** We measured distinctiveness as the inverse popularity of pages in each of the three choice sets. The popularity of a Facebook page was operationalized as the total number of users within the myPersonality dataset who followed that page (after filtering for US users for whom age, gender, and personality were recorded). For example, the page of Eminem was followed by 12,353 users whereas Edith Piaf was followed by only 69 users. Popularity scores for each of the three choice sets were calculated as the median popularity score of the 50 pages in their respective sets. We chose the median over the average to reduce the impact of extreme outliers. We then reverse-coded this score such that higher values indicated greater distinctiveness (i.e., lower popularity).

***Choice Diversity.*** We calculated choice diversity using two different metrics, which are largely uncorrelated (average correlation of $r$ = -0.06 across all three conditions) and therefore distinct. The first metric, topical diversity, is measured as entropy and captures how evenly spread a person's choices are across a set of categories. Specifically, it provides an estimate of how diverse a user's Facebook pages are when it comes to their associated categories (e.g., music, TV, business). We use normalized Shannon entropy as our main entropy metric, calculated as follows:

$$H_{norm} = \frac{\sum_{i=1}^{k} p_i \log_2 p_i}{\log_2 k}$$

With

$k$ = total number of page categories

$p_i$ = proportion (or probability) of any particular category (e.g. music)

The measure indicates how diverse the set of pages are relative to how diverse they could be (0 = no diversity with all pages falling under the same category, and 1 = maximum diversity with pages evenly spread across all categories). We use the normalized form of Shannon entropy to account for the fact that users have a different number of categories available in the first place. As an additional robustness check, we also calculate global evenness which compares the spread of pages to the overall number of categories in the sample (as opposed to the number of categories associated with a particular user).

The second metric, psychological interest diversity, "captures the extent to which individuals show diversity in the psychological profiles associated with their interests"[28] (see **Fig. 3** for a visual illustration of the operationalization). The measure has been used in prior research to reflect whether a person's Facebook pages are predominantly associated with a particular type of psychological profile (e.g., extraverted liberals) or a variety of different psychological profiles (e.g., extraverted liberals and introverted conservatives). We focus our measure of psychological interest diversity on the Big Five personality traits[29], which capture individual differences in the way that people think, feel and behave across a broad variety of domains[30]. Following the steps outlined in the original paper, the psychological interest diversity of the two choice sets is calculated with a focus on the Big Five personality traits using the following procedure (see **Fig. 9** for a visual illustration of the process):

(i) The personality profile of a Facebook page is estimated by averaging the personality scores of all US-based myPersonality participants for each of the Big Five traits (e.g., the average Extraversion score of all users following Lady Gaga's page).

(ii) For each user and Big Five personality trait, the diversity of each choice set is calculated as the standard deviation of the Facebook pages in each set (e.g., the standard deviation in Extraversion scores associated with the 50 Pages chosen for a user).

(iii) To arrive at a single measure of psychological interest diversity for each user, the diversity scores for each of the Big Five traits are averaged into a single diversity score for the entire choice set.

Higher scores on the interest diversity measure indicate that the choice set is more diverse and complex in the psychological makeup of its pages.

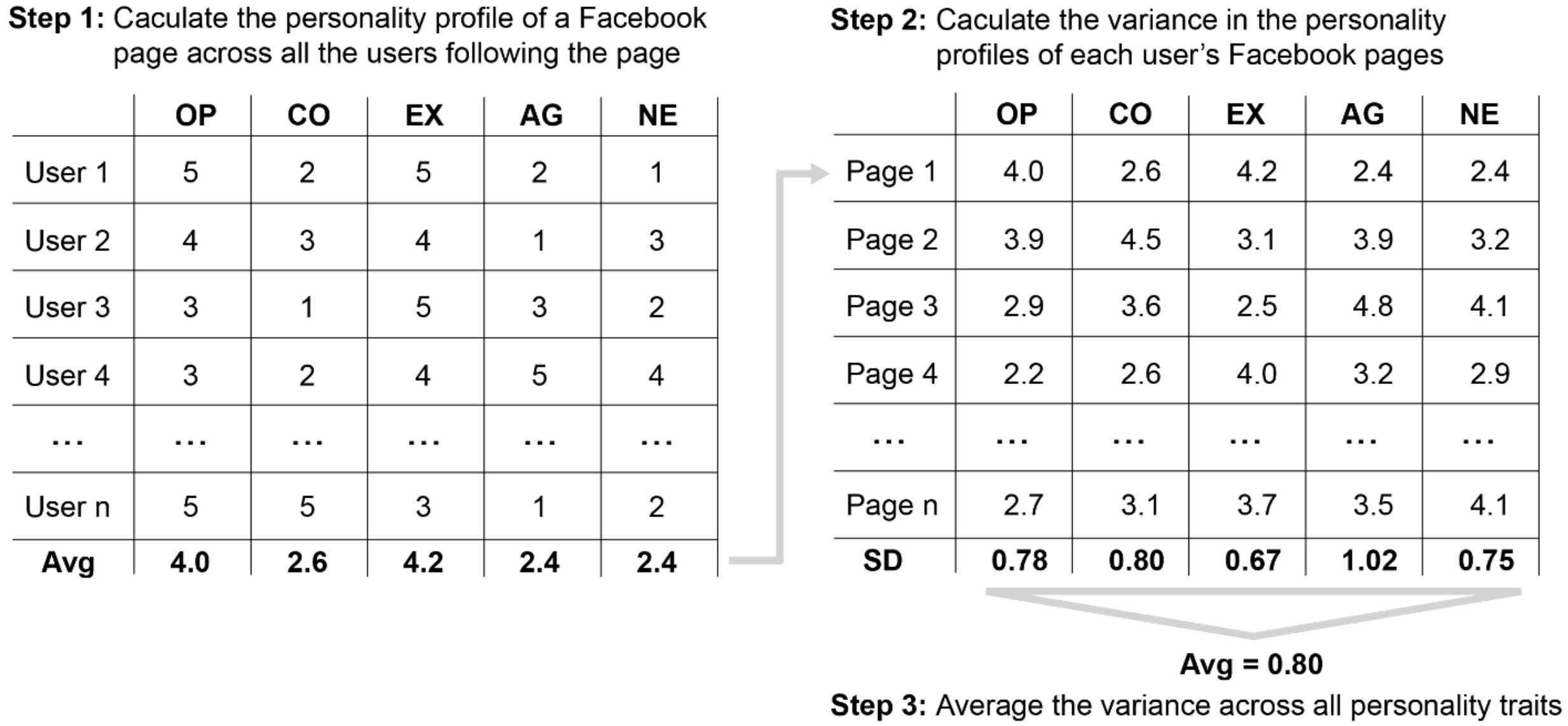

**Figure 9.** Operationalization of psychological interest diversity (adapted from Matz, 2017[28])

**Study 2.**

**Data.** We recruited 350 participants through CloudResearch Connect using a three-stage procedure. In the first stage, participants completed an initial screening study (compensated $0.50) in which they confirmed their willingness to download and upload their ChatGPT histories in a subsequent session. Several days later, participants were reinvited to upload their ChatGPT data; those who uploaded a valid file—defined as containing at least four weeks of usage data and a valid file structure—received $5. Uploaded chat histories were preprocessed, anonymized, and stripped of ChatGPT's responses, retaining only participants' messages. Because of context window constraints, chat histories were truncated to the most recent 227,000 tokens before being submitted via API to GPT-5.2, which generated a 20-line summary of each participant (for the prompt see **Fig. S3** in Appendix F in the SI). Participants were then reinvited to a final session, in which they completed the choice tasks described below and were compensated $7 upon completion.

Two participants were excluded for failing an attention check, resulting in a final analytic sample of 348 participants (*M* age = 33.2, *SD* = 10.8, range: 18–79; 189 male, 150 female, 9 other).

**Experimental Procedure.** In the third and final session, we deployed a within-person experimental design in which each participant selected between 10 and 15 items from a set of 150 movies, books, or podcasts, depending on the choice domain. Although each domain used a distinct visual theme (e.g., a Netflix-style browsing interface for movies), all three platforms followed an identical underlying logic and layout with items organized into horizontally scrollable category rows, an expandable item view, and a persistent collection panel for building one's choice set. Thus, the experience was consistent across domains apart from domain-specific content and surface-level styling. At the start of each domain, participants could also complete an optional interactive tutorial explaining how to browse items and build their collection.

Items were displayed under 6 categories for movies (Action, Sci-Fi, Drama, Comedy, Romance, and Documentary), 7 categories for books (Fiction, Sci-Fi/Fantasy, Non-Fiction, Thriller, Memoir, Self-Help, and Literary), and 6 categories for podcasts (True Crime, Society & Culture, Science & Technology, News & Politics, Business & Finance, and Health & Wellbeing). Participants selected items by adding them to a domain-specific collection panel (a "Watchlist" for movies, a "Basket" for books, and a "Playlist" for podcasts), from which items could be freely added or removed. They were required to reach a minimum of 10 items before submitting their final selection, and could not proceed before. Participants selected a mean of 12.09 (*SD* = 2.07), 11.37 (*SD* = 1.84), and 11.30 (*SD* = 1.80) items in the movies, books, and podcasts interfaces, respectively. This procedure resulted in a total of 12,097 choices across the three domains.

**Stimuli.** The item pools (150 movies, 150 books, and 150 podcasts) were assembled with the assistance of Claude Opus (Anthropic) which was prompted to generate a set of titles spanning six categories per domain (seven for books) that included a deliberate mix of mainstream and comparatively niche or lesser-known items, rather than exclusively popular titles. Each item was paired with a descriptive plot summary of approximately 120–160 words, generated using GPT-4o (OpenAI API). Because the item pools were curated rather than drawn from a naturally occurring catalog, popularity metadata was independently collected post-hoc for every item from external, real-world sources (see section on *Choice Distinctiveness* below). All stimuli can be accessed through the OSF repository.

**Display Ordering for Human Choices.** To generate the item ordering shown to participants in the generic and personalized recommendation conditions (see Procedure), we prompted GPT-5.2 (OpenAI API) with a numbered list of all 150 items in a domain (title, genre, and year or author where applicable) and asked it to return a ranked ordering of both the categories and the individual items within them. The generic prompt instructed the model to *"Order from most to least appealing to a general audience,"* whereas the personalized prompt included the participant's behavioral summary and instructed the model to *"Order from most to least relevant for this specific user."* The resulting rankings determined both the order in which categories were displayed and the ordering of items within each category. For prompts and output specifications, see **Fig. S4** in Appendix F in the SI.

**AI Agent Choice Prompting.** We generated AI-agent choice sets separately from the display-ordering manipulation described above, which only served as a more comprehensive human baseline. Unlike participants, the agents were not presented with one of the three ordered interfaces; instead, they received the item catalog directly as originally indexed and were asked to make their own selections. We varied these agents along two dimensions: the information available for personalization (none, domain-general, domain-specific, or both) and the selection process (batch or sequential). In every condition, GPT-5-mini received the catalog for the corresponding domain, with each item presented as its title followed by its description. The model was prompted to return only the titles of the selected items, which we matched back to the original catalog (see Appendix F, **Fig. S5-S6** in the SI for full prompts).

Across all conditions, the agents selected *n* items, where *n* corresponds to the number of items the participant had selected in the corresponding domain. For example, if a participant had chosen 12 movies, their agents would also choose 12 movies. The Generic Agent was given no information about the participant and asked to recommend *n* items from the full catalog. The personalized (domain-general) Agent received the same instructions prefaced by the participant's 20-line behavioral summary, with guidance to select items aligned with the person's inferred preferences.

The personalized (domain-specific) Agent instead received the items the participant had selected in that domain. Because these selections served as the information used to personalize the agent, it was asked to select *n* additional items, excluding those the participant had already selected. Finally, the personalized Hybrid Agent received both the behavioral summary and the participant's prior domain choices and were again prompted to select *n* additional items.

In batch conditions, the model selected all *n* items at once. In incremental conditions, all four personalization sources (Generic, Domain-General, Domain-Specific, Hybrid) were run using the same underlying logic, but with items selected sequentially, one at a time, over *n* steps, rather than all at once. At each step, the model recommended a single new item from the remaining candidate list, which was updated after every step to remove that item before the next prompt was issued. For the Domain-Specific and Hybrid conditions, each newly selected item was added to the context of prior choices provided in the following iteration's prompt.

**Measures** ***Choice Distinctiveness.*** As in Study 1, we measured distinctiveness as the reverse-coded popularity of items in each choice set. Popularity was operationalized separately for each domain. For movies, we queried the TMDB API for each title's popularity score—a composite score directly provided by TMDB, incorporating factors such as daily vote counts, view counts, "favorite" and "watchlist" additions, release date, total votes, and the previous day's score. For books, we scraped Goodreads for each title's ratings count; and for podcasts, we scraped Castbox.fm for each show's total play count. For each participant's (or agent's) set of selected items, we matched item titles back to the catalog using Levenshtein similarity, then calculated the popularity score as the median popularity of the matched items, consistent with our approach in Study 1, with lower popularity indicating higher levels of distinctiveness.

***Choice Diversity.*** We calculated choice diversity using two metrics. The first, topical diversity, was measured as normalized Shannon entropy and captures how evenly spread a person's (or agent's) choices are across the categories within a given domain, using the same formula reported in Study 1. Second, psychological diversity was measured as the average pairwise cosine distance between the description embeddings for the chosen items, generated using a pretrained sentence-embedding model (all-MiniLM-L6-v2). The specific operationalization of psychological diversity deviates from that used in Study 1, where the metric was calculated using Big Five personality profiles of each choice (i.e., the average personality profile of people following a Facebook page).

This deviation is because, unlike in Study 1, we did not have access to reliable Big Five profiles of the potential choices. Consequently, we decided to pursue a more data-driven approach that captured how semantically diverse a participant's (or agent's) selections were, without requiring those differences to map onto predefined personality dimensions. Although this approach is much broader than the narrow focus on Big Five personality traits, any psychological variation between items should be encoded in this measure. Higher scores indicate a more psychologically diverse choice set.

## References

1. Hill, K. I took a 'decision holiday' and put A.I in charge of my life. (2024).
2. Lai, V., Chen, C., Smith-Renner, A., Liao, Q. V. & Tan, C. Towards a Science of Human-AI Decision Making: An Overview of Design Space in Empirical Human-Subject Studies. in *2023 ACM Conference on Fairness Accountability and Transparency* 1369–1385 (ACM, Chicago IL USA, 2023). doi:10.1145/3593013.3594087.
3. Park, J. S. *et al.* Generative Agents: Interactive Simulacra of Human Behavior. in *Proceedings of the 36th Annual ACM Symposium on User Interface Software and Technology* 1–22 (ACM, San Francisco CA USA, 2023). doi:10.1145/3586183.3606763.
4. Yao, S. *et al.* React: Synergizing reasoning and acting in language models. in *International Conference on Learning Representations (ICLR)* (2023).
5. Talaifar, S. & Lowery, B. S. Freedom and Constraint in Digital Environments: Implications for the Self. *Perspect. Psychol. Sci.* **18**, 544–575 (2023).
6. Rajpurkar, P. *et al.* CheXaid: deep learning assistance for physician diagnosis of tuberculosis using chest x-rays in patients with HIV. *NPJ Digit. Med.* **3**, 115 (2020).
7. Bussmann, N., Giudici, P., Marinelli, D. & Papenbrock, J. Explainable Machine Learning in Credit Risk Management. *Comput. Econ.* **57**, 203–216 (2021).
8. Benjamin, D. M. *et al.* Hybrid forecasting of geopolitical events[†]. *AI Mag.* **44**, 112–128 (2023).
9. Brown, T. *et al.* Language models are few-shot learners. *Adv. Neural Inf. Process. Syst.* **33**, 1877–1901 (2020).
10. Doshi, A. R. & Hauser, O. P. Generative AI enhances individual creativity but reduces the collective diversity of novel content. *Sci. Adv.* **10**, eadn5290 (2024).

11. Moon, K., Green, A. & Kushlev, K. Homogenizing Effect of Large Language Model (LLM) on Creative Diversity: An Empirical Comparison. (2024).
12. Anderson, B. R., Shah, J. H. & Kreminski, M. Homogenization Effects of Large Language Models on Human Creative Ideation. in *Creativity and Cognition* 413–425 (ACM, Chicago IL USA, 2024). doi:10.1145/3635636.3656204.
13. Sourati, Z. *et al.* The Shrinking Landscape of Linguistic Diversity in the Age of Large Language Models. Preprint at https://doi.org/10.48550/arXiv.2502.11266 (2025).
14. Padmakumar, V. & He, H. Does writing with language models reduce content diversity? in *International Conference on Learning Representations* vol. 2024 642–669 (2024).
15. Goethals, S. & Rhue, L. One world, one opinion? The superstar effect in LLM responses. in *Proceedings of the 3rd Workshop on Cross-Cultural Considerations in NLP (C3NLP 2025)* 89–107 (2025).
16. Shur-Ofry, M., Horowitz-Amsalem, B., Belinkov, Y. & Rahamim, A. Growing a tail: Increasing output diversity in large language models. *Mach. Learn. Appl.* 100951 (2026).
17. Miller, M. E. & Spatz, E. A unified view of a human digital twin. *Hum.-Intell. Syst. Integr.* **4**, 23–33 (2022).
18. Peters, H. & Matz, S. C. Large language models can infer psychological dispositions of social media users. *PNAS Nexus* **3**, pgae231 (2024).
19. Goethals, S., Luther, J. & Matz, S. Words reveal wants: How well can simple LLM-based AI agents replicate people's choices based on their social media posts. in *Adjunct Proceedings of the 33rd ACM Conference on User Modeling, Adaptation and Personalization* 126–131 (ACM, New York City USA, 2025). doi:10.1145/3708319.3733689.

20. Kosinski, M., Matz, S. C., Gosling, S. D., Popov, V. & Stillwell, D. Facebook as a research tool for the social sciences: Opportunities, challenges, ethical considerations, and practical guidelines. *Am. Psychol.* **70**, 543 (2015).

21. Nguyen, T. T., Hui, P.-M., Harper, F. M., Terveen, L. & Konstan, J. A. Exploring the filter bubble: the effect of using recommender systems on content diversity. in *Proceedings of the 23rd international conference on World wide web* 677–686 (ACM, Seoul Korea, 2014). doi:10.1145/2566486.2568012.

22. Padmakumar, V. & He, H. Does Writing with Language Models Reduce Content Diversity? Preprint at https://doi.org/10.48550/arXiv.2309.05196 (2024).

23. Klingbeil, A., Grützner, C. & Schreck, P. Trust and reliance on AI—An experimental study on the extent and costs of overreliance on AI. *Comput. Hum. Behav.* **160**, 108352 (2024).

24. Bommasani, R. *et al.* On the Opportunities and Risks of Foundation Models. Preprint at https://doi.org/10.48550/arXiv.2108.07258 (2022).

25. Wenger, E. & Kenett, Y. N. Large language models are homogeneously creative. *PNAS Nexus* **5**, pgag042 (2026).

26. Ziegler, C.-N., McNee, S. M., Konstan, J. A. & Lausen, G. Improving recommendation lists through topic diversification. in *Proceedings of the 14th international conference on World Wide Web - WWW '05* 22 (ACM Press, Chiba, Japan, 2005). doi:10.1145/1060745.1060754.

27. McNee, S. M., Riedl, J. & Konstan, J. A. Being accurate is not enough: how accuracy metrics have hurt recommender systems. in *CHI '06 Extended Abstracts on Human Factors in Computing Systems* 1097–1101 (ACM, Montréal Québec Canada, 2006). doi:10.1145/1125451.1125659.

28. Matz, S. C. Personal echo chambers: Openness-to-experience is linked to higher levels of psychological interest diversity in large-scale behavioral data. *J. Pers. Soc. Psychol.* **121**, 1284 (2021).

29. John, O. P. & Srivastava, S. The Big-Five trait taxonomy: History, measurement, and theoretical perspectives. http://www.personality-project.org/revelle/syllabi/classreadings/john.pdf (1999).

30. Ozer, D. J. & Benet-Martínez, V. Personality and the Prediction of Consequential Outcomes. *Annu. Rev. Psychol.* **57**, 401–421 (2006).

# Supplementary Materials

## **Appendix A** – Visualization of the analytical procedure applied in Study 1.

**Figure S1.** Analytical procedure for a hypothetical participant X.

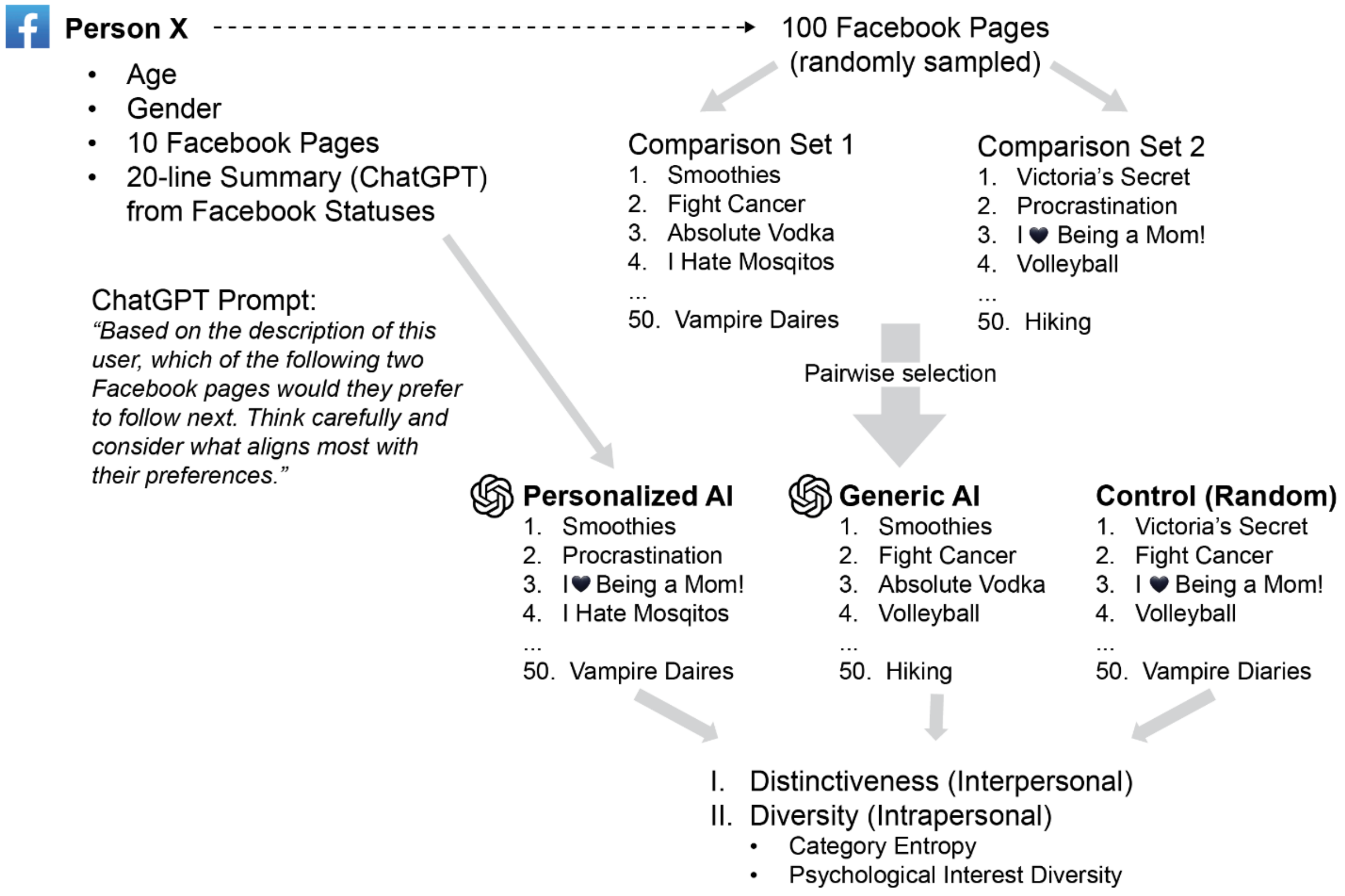

**Appendix B** – Pairwise contrasts of the different experimental conditions in Study 2 for each of the three outcome variables.

**Table S1.** Pairwise contrasts from a linear regression analysis predicting distinctiveness from the experimental condition

| Contrast | Estimate | SE | *df* | *t* | *p* |
|---|---|---|---|---|---|
| (1) Random - (2) Generic AI | 203,653.9 | 143,909.2 | 1,039 | 1.42 | 0.333 |
| (1) Random - (3) Personalized AI | 365,117.3 | 143,903.9 | 1,039 | 2.54 | 0.030 |
| (2) Generic AI - (3) Personalized AI | 161,463.4 | 143,909.2 | 1,039 | 1.12 | 0.501 |

Note. Estimates are based on estimated marginal means. *p* values are Tukey-adjusted.

**Table S2.** Pairwise contrasts from a linear regression analysis predicting psychological diversity (cosine distance) from the experimental condition

| Contrast | Estimate | SE | *df* | *t* | *p* |
|---|---|---|---|---|---|
| (1) Random - (2) Generic AI | 0 | 0 | 1,039 | 0.92 | 0.629 |
| (1) Random - (3) Personalized AI | 0 | 0 | 1,039 | 1.22 | 0.442 |
| (2) Generic AI - (3) Personalized AI | 0 | 0 | 1,039 | 0.30 | 0.951 |

Note. Estimates are based on estimated marginal means. *p* values are Tukey-adjusted.

**Table S3.** Pairwise contrasts from a linear regression analysis predicting topical diversity (entropy) from the experimental condition

| Contrast | Estimate | SE | *df* | *t* | *p* |
|---|---|---|---|---|---|
| (1) Random - (2) Generic AI | 0.00 | 0.01 | 1,039 | -0.25 | 0.966 |
| (1) Random - (3) Personalized AI | -0.02 | 0.01 | 1,039 | -1.40 | 0.342 |
| (2) Generic AI - (3) Personalized AI | -0.01 | 0.01 | 1,039 | -1.15 | 0.484 |

Note. Estimates are based on estimated marginal means. *p* values are Tukey-adjusted.

**Appendix C** – Linear Mixed-Effect Models predicting all three outcomes from different agentic conditions (with Random selection as the reference group). Observations are cross nested in the Domain (i.e., movie, book, podcast) and participant ID. $N = 10{,}440$ observations from

**Table S4.** Linear Mixed-Effects Model predicting Distinctiveness

| Predictor | Estimate | SE | *t* | *p* | 95% CI |
|---|---|---|---|---|---|
| Human Choice | -0.63 | 0.03 | -22.38 | < .001 | [-0.69, -0.58] |
| Generic (Batch) | -2.05 | 0.03 | -72.71 | < .001 | [-2.1, -1.99] |
| Personalized (Domain-general\|Batch) | -0.15 | 0.03 | -5.49 | < .001 | [-0.21, -0.1] |
| Personalized (Domain-specific\|Batch) | -0.76 | 0.03 | -26.88 | < .001 | [-0.81, -0.7] |
| Personalized (Hybrid\|Batch) | -0.19 | 0.03 | -6.57 | < .001 | [-0.24, -0.13] |
| Generic (Sequential) | -2.20 | 0.03 | -78.03 | < .001 | [-2.25, -2.14] |
| Personalized (Domain-general\|Sequential) | -0.26 | 0.03 | -9.24 | < .001 | [-0.32, -0.21] |
| Personalized (Domain-specific\|Sequential) | -0.31 | 0.03 | -11.13 | < .001 | [-0.37, -0.26] |
| Personalized (Hybrid\|Sequential) | -0.18 | 0.03 | -6.23 | < .001 | [-0.23, -0.12] |

Note. CI = confidence interval. The outcome variable was *z*-standardized within Task.

**Table S5.** Linear Mixed-Effects Model Predicting Diversity (Psychological)

| Predictor | Estimate | SE | *t* | *p* | 95% CI |
|---|---|---|---|---|---|
| Human Choice | -0.46 | 0.04 | -11.53 | < .001 | [-0.53, -0.38] |
| Generic (Batch) | -0.20 | 0.04 | -5.00 | < .001 | [-0.27, -0.12] |
| Personalized (Domain-general\|Batch) | -0.62 | 0.04 | -15.74 | < .001 | [-0.7, -0.54] |
| Personalized (Domain-specific\|Batch) | -0.72 | 0.04 | -18.24 | < .001 | [-0.8, -0.64] |
| Personalized (Hybrid\|Batch) | -0.43 | 0.04 | -10.93 | < .001 | [-0.51, -0.35] |
| Generic (Sequential) | -0.59 | 0.04 | -14.94 | < .001 | [-0.67, -0.51] |
| Personalized (Domain-general\|Sequential) | -0.84 | 0.04 | -21.17 | < .001 | [-0.91, -0.76] |
| Personalized (Domain-specific\|Sequential) | -1.39 | 0.04 | -35.28 | < .001 | [-1.47, -1.32] |

| Predictor | Estimate | SE | *t* | *p* | 95% CI |
|---|---|---|---|---|---|
| Personalized (Hybrid\|Sequential) | -1.04 | 0.04 | -26.27 | < .001 | [-1.11, -0.96] |

Note. CI = confidence interval. The outcome variable was *z*-standardized within Task.

**Table S6.** Linear Mixed-Effects Model Predicting Diversity (Topical)

| Predictor | Estimate | SE | *t* | *p* | 95% CI |
|---|---|---|---|---|---|
| Human Choice | -0.36 | 0.04 | -9.44 | < .001 | [-0.43, -0.28] |
| Generic (Batch) | -0.10 | 0.04 | -2.74 | .006 | [-0.18, -0.03] |
| Personalized (Domain-general\|Batch) | -0.37 | 0.04 | -9.88 | < .001 | [-0.45, -0.3] |
| Personalized (Domain-specific\|Batch) | -0.39 | 0.04 | -10.38 | < .001 | [-0.47, -0.32] |
| Personalized (Hybrid\|Batch) | -0.38 | 0.04 | -9.94 | < .001 | [-0.45, -0.3] |
| Generic (Sequential) | -1.05 | 0.04 | -27.68 | < .001 | [-1.12, -0.97] |
| Personalized (Domain-general\|Sequential) | -0.49 | 0.04 | -13.04 | < .001 | [-0.57, -0.42] |
| Personalized (Domain-specific\|Sequential) | -1.72 | 0.04 | -45.44 | < .001 | [-1.79, -1.64] |
| Personalized (Hybrid\|Sequential) | -1.08 | 0.04 | -28.49 | < .001 | [-1.15, -1] |

Note. CI = confidence interval. The outcome variable was *z*-standardized within Task.

**Appendix D** – Breakdown of the effects of different design choices on distinctiveness and diversity across all three domains (see Fig. S7 in the main manuscript for a summary of findings collapsed across domains).

**Figure S2.** Average scores (and standard errors) for choice distinctiveness (row 1) and choice diversity outcomes (rows 2-3), separately for all choice domains. All scores were *z*-standardized within outcome and domain before being averaged

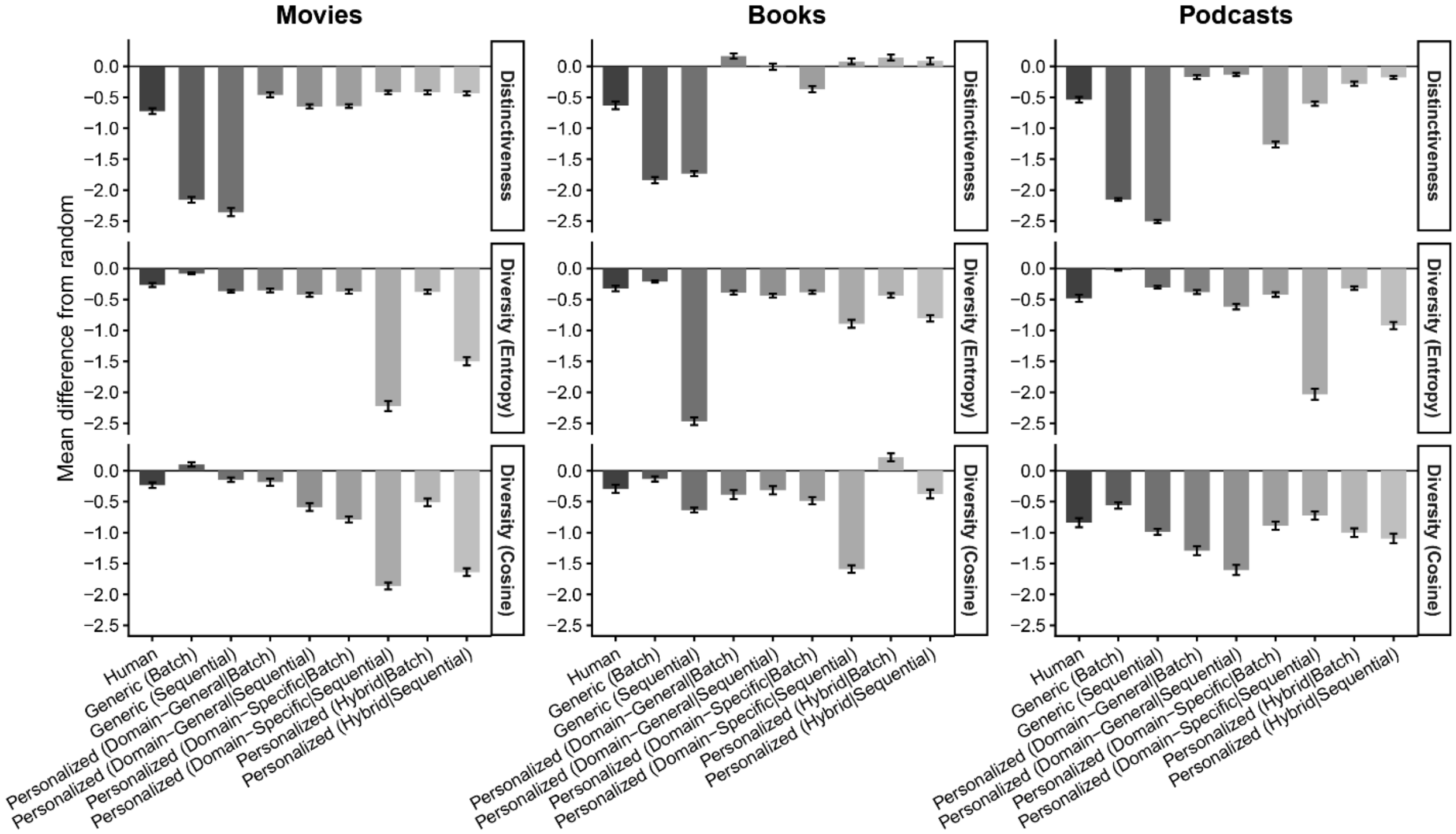

.

**Appendix E** - Pairwise contrasts of the different AI conditions (+ human and random) in Study 2 for each of the three outcome variables.

**Table S7.** Pairwise contrasts from a linear mixed-effects regression analysis (see Table S4) predicting distinctiveness from the AI agent condition

| Contrast | Estimate | SE | *z* | *p* | *d* |
|---|---|---|---|---|---|
| Random - Human | 0.63 | 0.03 | 22.38 | < .001 | 0.98 |
| Random – Generic (Batch) | 2.05 | 0.03 | 72.71 | < .001 | 3.18 |
| Random - Personalized (Domain-general\|Batch) | 0.15 | 0.03 | 5.49 | < .001 | 0.24 |
| Random - Personalized (Domain-specific\|Batch) | 0.76 | 0.03 | 26.88 | < .001 | 1.18 |
| Random - Personalized (Hybrid\|Batch) | 0.19 | 0.03 | 6.57 | < .001 | 0.29 |
| Random - Generic (Sequential) | 2.20 | 0.03 | 78.03 | < .001 | 3.42 |
| Random - Personalized (Domain-general\|Sequential) | 0.26 | 0.03 | 9.24 | < .001 | 0.40 |
| Random - Personalized (Domain-specific\| Sequential) | 0.31 | 0.03 | 11.13 | < .001 | 0.49 |
| Random - Personalized (Hybrid\| Sequential) | 0.18 | 0.03 | 6.23 | < .001 | 0.27 |
| Human - Generic (Batch) | 1.42 | 0.03 | 50.33 | < .001 | 2.20 |
| Human - Personalized (Domain-general\|Batch) | -0.48 | 0.03 | -16.89 | < .001 | -0.74 |
| Human - Personalized (Domain-specific\|Batch) | 0.13 | 0.03 | 4.50 | < .001 | 0.20 |
| Human - Personalized (Hybrid\|Batch) | -0.45 | 0.03 | -15.81 | < .001 | -0.69 |
| Human - Generic (Sequential) | 1.57 | 0.03 | 55.65 | < .001 | 2.44 |
| Human - Personalized (Domain-general\| Sequential) | -0.37 | 0.03 | -13.14 | < .001 | -0.57 |
| Human - Personalized (Domain-specific\| Sequential) | -0.32 | 0.03 | -11.25 | < .001 | -0.49 |
| Human - Personalized (Hybrid\| Sequential) | -0.46 | 0.03 | -16.15 | < .001 | -0.71 |
| Generic (Batch) - Personalized (Domain-general\|Batch) | -1.89 | 0.03 | -67.22 | < .001 | -2.94 |
| Generic (Batch) - Personalized (Domain-specific\|Batch) | -1.29 | 0.03 | -45.83 | < .001 | -2.01 |
| Generic (Batch) - Personalized (Hybrid\|Batch) | -1.86 | 0.03 | -66.14 | < .001 | -2.89 |
| Generic (Batch) - Generic (Sequential) | 0.15 | 0.03 | 5.32 | < .001 | 0.23 |
| Generic (Batch) - Personalized (Domain-general\| Sequential) | -1.79 | 0.03 | -63.46 | < .001 | -2.78 |
| Generic (Batch) - Personalized (Domain-specific\| Sequential) | -1.74 | 0.03 | -61.58 | < .001 | -2.70 |
| Generic (Batch) - Personalized (Hybrid\| Sequential) | -1.87 | 0.03 | -66.48 | < .001 | -2.91 |
| Personalized (Domain-general\|Batch) - Personalized (Domain-specific\|Batch) | 0.60 | 0.03 | 21.39 | < .001 | 0.94 |

| Contrast | Estimate | SE | *z* | *p* | *d* |
|---|---|---|---|---|---|
| Personalized (Domain-general\|Batch) - Personalized (Hybrid\|Batch) | 0.03 | 0.03 | 1.08 | 0.987 | 0.05 |
| Personalized (Domain-general\|Batch) - Generic (Sequential) | 2.04 | 0.03 | 72.54 | < .001 | 3.17 |
| Personalized (Domain-general\|Batch) - Personalized (Domain-general\| Sequential) | 0.11 | 0.03 | 3.75 | 0.007 | 0.16 |
| Personalized (Domain-general\|Batch) - Personalized (Domain-specific\| Sequential) | 0.16 | 0.03 | 5.64 | < .001 | 0.25 |
| Personalized (Domain-general\|Batch) - Personalized (Hybrid\| Sequential) | 0.02 | 0.03 | 0.74 | 0.999 | 0.03 |
| Personalized (Domain-specific\|Batch) - Personalized (Hybrid\|Batch) | -0.57 | 0.03 | -20.31 | < .001 | -0.89 |
| Personalized (Domain-specific\|Batch) - Generic (Sequential) | 1.44 | 0.03 | 51.15 | < .001 | 2.24 |
| Personalized (Domain-specific\|Batch) - Personalized (Domain-general\| Sequential) | -0.50 | 0.03 | -17.64 | < .001 | -0.77 |
| Personalized (Domain-specific\|Batch) - Personalized (Domain-specific\| Sequential) | -0.44 | 0.03 | -15.75 | < .001 | -0.69 |
| Personalized (Domain-specific\|Batch) - Personalized (Hybrid\| Sequential) | -0.58 | 0.03 | -20.65 | < .001 | -0.90 |
| Personalized (Hybrid\|Batch) - Generic (Sequential) | 2.01 | 0.03 | 71.46 | < .001 | 3.13 |
| Personalized (Hybrid\|Batch) - (Personalized (Domain-general\| Sequential) | 0.08 | 0.03 | 2.68 | 0.184 | 0.12 |
| Personalized (Hybrid\|Batch) - Personalized (Domain-specific\| Sequential) | 0.13 | 0.03 | 4.56 | < .001 | 0.20 |
| Personalized (Hybrid\|Batch) - Personalized (Hybrid\| Sequential) | -0.01 | 0.03 | -0.34 | 1.000 | -0.01 |
| Generic (Sequential) - Personalized (Domain-general\| Sequential) | -1.94 | 0.03 | -68.78 | < .001 | -3.01 |
| Generic (Sequential) - Personalized (Domain-specific\| Sequential) | -1.89 | 0.03 | -66.90 | < .001 | -2.93 |
| Generic (Sequential) - Personalized (Hybrid\| Sequential) | -2.02 | 0.03 | -71.80 | < .001 | -3.14 |
| (Personalized (Domain-general\| Sequential)- (Personalized (Domain-specific\| Sequential) | 0.05 | 0.03 | 1.89 | 0.679 | 0.08 |
| (Personalized (Domain-general\| Sequential)- (Personalized (Hybrid\| Sequential) | -0.08 | 0.03 | -3.01 | 0.077 | -0.13 |
| (Personalized (Domain-specific\| Sequential)- (Personalized (Hybrid\| Sequential) | -0.14 | 0.03 | -4.90 | < .001 | -0.21 |

Note. Estimates are based on estimated marginal means. *p* values are Tukey-adjusted.

**Table S8.** Pairwise contrasts from a linear mixed-effects regression analysis (see Table S5) predicting psychological diversity from the AI agent condition

| Contrast | Estimate | SE | *z* | *p* | *d* |
|---|---|---|---|---|---|
| Random - Human | 0.46 | 0.04 | 11.53 | < .001 | 0.50 |
| Random – Generic (Batch) | 0.20 | 0.04 | 5.00 | < .001 | 0.22 |
| Random - Personalized (Domain-general\|Batch) | 0.62 | 0.04 | 15.74 | < .001 | 0.69 |
| Random - Personalized (Domain-specific\|Batch) | 0.72 | 0.04 | 18.24 | < .001 | 0.80 |
| Random - Personalized (Hybrid\|Batch) | 0.43 | 0.04 | 10.93 | < .001 | 0.48 |
| Random - Generic (Sequential) | 0.59 | 0.04 | 14.94 | < .001 | 0.65 |
| Random - Personalized (Domain-general\|Sequential) | 0.84 | 0.04 | 21.17 | < .001 | 0.93 |
| Random - Personalized (Domain-specific\| Sequential) | 1.39 | 0.04 | 35.28 | < .001 | 1.54 |
| Random - Personalized (Hybrid\| Sequential) | 1.04 | 0.04 | 26.27 | < .001 | 1.15 |
| Human - Generic (Batch) | -0.26 | 0.04 | -6.53 | < .001 | -0.29 |
| Human - Personalized (Domain-general\|Batch) | 0.17 | 0.04 | 4.22 | 0.001 | 0.18 |
| Human - Personalized (Domain-specific\|Batch) | 0.27 | 0.04 | 6.71 | < .001 | 0.29 |
| Human - Personalized (Hybrid\|Batch) | -0.02 | 0.04 | -0.60 | 1.000 | -0.03 |
| Human - Generic (Sequential) | 0.13 | 0.04 | 3.41 | 0.023 | 0.15 |
| Human - Personalized (Domain-general\| Sequential) | 0.38 | 0.04 | 9.64 | < .001 | 0.42 |
| Human - Personalized (Domain-specific\| Sequential) | 0.94 | 0.04 | 23.75 | < .001 | 1.04 |
| Human - Personalized (Hybrid\| Sequential) | 0.58 | 0.04 | 14.74 | < .001 | 0.65 |
| Generic (Batch) - Personalized (Domain-general\|Batch) | 0.42 | 0.04 | 10.75 | < .001 | 0.47 |
| Generic (Batch) - Personalized (Domain-specific\|Batch) | 0.52 | 0.04 | 13.24 | < .001 | 0.58 |
| Generic (Batch) - Personalized (Hybrid\|Batch) | 0.23 | 0.04 | 5.93 | < .001 | 0.26 |
| Generic (Batch) - Generic (Sequential) | 0.39 | 0.04 | 9.94 | < .001 | 0.44 |
| Generic (Batch) - Personalized (Domain-general\| Sequential) | 0.64 | 0.04 | 16.17 | < .001 | 0.71 |
| Generic (Batch) - Personalized (Domain-specific\| Sequential) | 1.20 | 0.04 | 30.28 | < .001 | 1.33 |
| Generic (Batch) - Personalized (Hybrid\| Sequential) | 0.84 | 0.04 | 21.27 | < .001 | 0.93 |
| Personalized (Domain-general\|Batch) - Personalized (Domain-specific\|Batch) | 0.10 | 0.04 | 2.49 | 0.272 | 0.11 |
| Personalized (Domain-general\|Batch) - Personalized (Hybrid\|Batch) | -0.19 | 0.04 | -4.82 | < .001 | -0.21 |
| Personalized (Domain-general\|Batch) - Generic (Sequential) | -0.03 | 0.04 | -0.80 | 0.999 | -0.04 |

| Contrast | Estimate | SE | *z* | *p* | *d* |
|---|---|---|---|---|---|
| Personalized (Domain-general\|Batch) - Personalized (Domain-general\| Sequential) | 0.21 | 0.04 | 5.42 | < .001 | 0.24 |
| Personalized (Domain-general\|Batch) - Personalized (Domain-specific\| Sequential) | 0.77 | 0.04 | 19.53 | < .001 | 0.85 |
| Personalized (Domain-general\|Batch) - Personalized (Hybrid\| Sequential) | 0.42 | 0.04 | 10.53 | < .001 | 0.46 |
| Personalized (Domain-specific\|Batch) - Personalized (Hybrid\|Batch) | -0.29 | 0.04 | -7.31 | < .001 | -0.32 |
| Personalized (Domain-specific\|Batch) - Generic (Sequential) | -0.13 | 0.04 | -3.30 | 0.033 | -0.14 |
| Personalized (Domain-specific\|Batch) - Personalized (Domain-general\| Sequential) | 0.12 | 0.04 | 2.93 | 0.098 | 0.13 |
| Personalized (Domain-specific\|Batch) - Personalized (Domain-specific\| Sequential) | 0.67 | 0.04 | 17.04 | < .001 | 0.75 |
| Personalized (Domain-specific\|Batch) - Personalized (Hybrid\| Sequential) | 0.32 | 0.04 | 8.03 | < .001 | 0.35 |
| Personalized (Hybrid\|Batch) - Generic (Sequential) | 0.16 | 0.04 | 4.01 | 0.002 | 0.18 |
| Personalized (Hybrid\|Batch) - (Personalized (Domain-general\| Sequential) | 0.40 | 0.04 | 10.24 | < .001 | 0.45 |
| Personalized (Hybrid\|Batch) - Personalized (Domain-specific\| Sequential) | 0.96 | 0.04 | 24.35 | < .001 | 1.07 |
| Personalized (Hybrid\|Batch) - Personalized (Hybrid\| Sequential) | 0.61 | 0.04 | 15.35 | < .001 | 0.67 |
| Generic (Sequential) - Personalized (Domain-general\| Sequential) | 0.25 | 0.04 | 6.23 | < .001 | 0.27 |
| Generic (Sequential) - Personalized (Domain-specific\| Sequential) | 0.80 | 0.04 | 20.34 | < .001 | 0.89 |
| Generic (Sequential) - Personalized (Hybrid\| Sequential) | 0.45 | 0.04 | 11.33 | < .001 | 0.50 |
| (Personalized (Domain-general\| Sequential)- (Personalized (Domain-specific\| Sequential) | 0.56 | 0.04 | 14.11 | < .001 | 0.62 |
| (Personalized (Domain-general\| Sequential)- (Personalized (Hybrid\| Sequential) | 0.20 | 0.04 | 5.11 | < .001 | 0.22 |
| (Personalized (Domain-specific\| Sequential)- (Personalized (Hybrid\| Sequential) | -0.36 | 0.04 | -9.01 | < .001 | -0.39 |

Note. Estimates are based on estimated marginal means. *p* values are Tukey-adjusted.

**Table S9.** Pairwise contrasts from a linear mixed-effects regression analysis (see Table S6) predicting topical diversity from the AI agent condition

| Contrast | Estimate | SE | *z* | *p* | *d* |
|---|---|---|---|---|---|
| Random - Human | 0.36 | 0.04 | 9.44 | < .001 | 0.41 |
| Random – Generic (Batch) | 0.10 | 0.04 | 2.74 | 0.156 | 0.12 |
| Random - Personalized (Domain-general\|Batch) | 0.37 | 0.04 | 9.88 | < .001 | 0.43 |
| Random - Personalized (Domain-specific\|Batch) | 0.39 | 0.04 | 10.38 | < .001 | 0.45 |
| Random - Personalized (Hybrid\|Batch) | 0.38 | 0.04 | 9.94 | < .001 | 0.44 |
| Random - Generic (Sequential) | 1.05 | 0.04 | 27.68 | < .001 | 1.21 |
| Random - Personalized (Domain-general\|Sequential) | 0.49 | 0.04 | 13.04 | < .001 | 0.57 |
| Random - Personalized (Domain-specific\| Sequential) | 1.72 | 0.04 | 45.44 | < .001 | 1.99 |
| Random - Personalized (Hybrid\| Sequential) | 1.08 | 0.04 | 28.49 | < .001 | 1.25 |
| Human - Generic (Batch) | -0.25 | 0.04 | -6.69 | < .001 | -0.29 |
| Human - Personalized (Domain-general\|Batch) | 0.02 | 0.04 | 0.44 | 1.000 | 0.02 |
| Human - Personalized (Domain-specific\|Batch) | 0.04 | 0.04 | 0.94 | 0.995 | 0.04 |
| Human - Personalized (Hybrid\|Batch) | 0.02 | 0.04 | 0.50 | 1.000 | 0.02 |
| Human - Generic (Sequential) | 0.69 | 0.04 | 18.25 | < .001 | 0.80 |
| Human - Personalized (Domain-general\| Sequential) | 0.14 | 0.04 | 3.60 | 0.012 | 0.16 |
| Human - Personalized (Domain-specific\| Sequential) | 1.36 | 0.04 | 36.00 | < .001 | 1.58 |
| Human - Personalized (Hybrid\| Sequential) | 0.72 | 0.04 | 19.05 | < .001 | 0.83 |
| Generic (Batch) - Personalized (Domain-general\|Batch) | 0.27 | 0.04 | 7.13 | < .001 | 0.31 |
| Generic (Batch) - Personalized (Domain-specific\|Batch) | 0.29 | 0.04 | 7.63 | < .001 | 0.33 |
| Generic (Batch) - Personalized (Hybrid\|Batch) | 0.27 | 0.04 | 7.20 | < .001 | 0.31 |
| Generic (Batch) - Generic (Sequential) | 0.94 | 0.04 | 24.94 | < .001 | 1.09 |
| Generic (Batch) - Personalized (Domain-general\| Sequential) | 0.39 | 0.04 | 10.29 | < .001 | 0.45 |
| Generic (Batch) - Personalized (Domain-specific\| Sequential) | 1.61 | 0.04 | 42.70 | < .001 | 1.87 |
| Generic (Batch) - Personalized (Hybrid\| Sequential) | 0.97 | 0.04 | 25.75 | < .001 | 1.13 |
| Personalized (Domain-general\|Batch) - Personalized (Domain-specific\|Batch) | 0.02 | 0.04 | 0.50 | 1.000 | 0.02 |
| Personalized (Domain-general\|Batch) - Personalized (Hybrid\|Batch) | 0.00 | 0.04 | 0.06 | 1.000 | 0.00 |
| Personalized (Domain-general\|Batch) - Generic (Sequential) | 0.67 | 0.04 | 17.81 | < .001 | 0.78 |

| Contrast | Estimate | SE | *z* | *p* | *d* |
|---|---|---|---|---|---|
| Personalized (Domain-general\|Batch) - Personalized (Domain-general\| Sequential) | 0.12 | 0.04 | 3.16 | 0.051 | 0.14 |
| Personalized (Domain-general\|Batch) - Personalized (Domain-specific\| Sequential) | 1.34 | 0.04 | 35.57 | < .001 | 1.56 |
| Personalized (Domain-general\|Batch) - Personalized (Hybrid\| Sequential) | 0.70 | 0.04 | 18.62 | < .001 | 0.81 |
| Personalized (Domain-specific\|Batch) - Personalized (Hybrid\|Batch) | -0.02 | 0.04 | -0.44 | 1.000 | -0.02 |
| Personalized (Domain-specific\|Batch) - Generic (Sequential) | 0.65 | 0.04 | 17.31 | < .001 | 0.76 |
| Personalized (Domain-specific\|Batch) - Personalized (Domain-general\| Sequential) | 0.10 | 0.04 | 2.66 | 0.190 | 0.12 |
| Personalized (Domain-specific\|Batch) - Personalized (Domain-specific\| Sequential) | 1.32 | 0.04 | 35.07 | < .001 | 1.53 |
| Personalized (Domain-specific\|Batch) - Personalized (Hybrid\| Sequential) | 0.68 | 0.04 | 18.12 | < .001 | 0.79 |
| Personalized (Hybrid\|Batch) - Generic (Sequential) | 0.67 | 0.04 | 17.74 | < .001 | 0.78 |
| Personalized (Hybrid\|Batch) - (Personalized (Domain-general\| Sequential) | 0.12 | 0.04 | 3.10 | 0.061 | 0.14 |
| Personalized (Hybrid\|Batch) - Personalized (Domain-specific\| Sequential) | 1.34 | 0.04 | 35.50 | < .001 | 1.55 |
| Personalized (Hybrid\|Batch) - Personalized (Hybrid\| Sequential) | 0.70 | 0.04 | 18.55 | < .001 | 0.81 |
| Generic (Sequential) - Personalized (Domain-general\| Sequential) | -0.55 | 0.04 | -14.65 | < .001 | -0.64 |
| Generic (Sequential) - Personalized (Domain-specific\| Sequential) | 0.67 | 0.04 | 17.76 | < .001 | 0.78 |
| Generic (Sequential) - Personalized (Hybrid\| Sequential) | 0.03 | 0.04 | 0.81 | 0.999 | 0.04 |
| (Personalized (Domain-general\| Sequential)- (Personalized (Domain-specific\| Sequential) | 1.22 | 0.04 | 32.41 | < .001 | 1.42 |
| (Personalized (Domain-general\| Sequential)- (Personalized (Hybrid\| Sequential) | 0.58 | 0.04 | 15.46 | < .001 | 0.68 |
| (Personalized (Domain-specific\| Sequential)- (Personalized (Hybrid\| Sequential) | -0.64 | 0.04 | -16.95 | < .001 | -0.74 |

Note. Estimates are based on estimated marginal means. *p* values are Tukey-adjusted.

**Appendix F** – Prompts used in Study 2

**Figure S3.** Participant Summary

```
# Role #
You are a behavioral scientist and expert profiler specializing in psychographic
summaries based on conversational data.

# Task #
You are given a ChatGPT conversation history that contains only the user’s
messages.
Your task is to infer a detailed psychographic description of the individual
behind the messages.

Base your description on an integrated analysis of:
- WHAT the person chooses to talk about (recurring topics, themes, interests,
concerns, omissions)
- HOW the person communicates (tone, structure, emotional expression, control,
iteration, humor, openness)

# Output Format #
- Your response must ONLY contain the description.
- Do NOT include any preambles, greetings, sign-offs, or any other text before
or after the description itself.
- Send your response in English.
- Each line must capture a distinct, non-redundant insight grounded in
observable behavior or content.
- The description must not exceed 20 lines.
```

**Figure S4.** Display Ordering in the Human Choice conditions

Generic

```
You are part of a recommendation system that helps users find the best and most
relevant {type}. Below is a set of {type}:

{list}

Respond with exactly TWO lines and nothing else:
Line 1: a JSON array of the {n_categories} category/genre names ordered from
most to least relevant. The categories are: {category_list}
Line 2: a JSON array of all {n_items} original 1-based item numbers in
recommended order.
Example format:
["CategoryA","CategoryB"]
[3,1,5,...]
```

Personalized

```
You are part of a recommendation system that helps users find the best and most
relevant {type}. Below is a set of {type}:

{list}

Below is the summary profile of the user you are recommending {type} for. Please
tailor your recommendations to the specific preferences of the user to make them
as relevant and useful as possible for them.
```

```
User profile: {profile}.

Respond with exactly TWO lines and nothing else:
Line 1: a JSON array of the {n_categories} category/genre names ordered from
most to least relevant. The categories are: {category_list}
Line 2: a JSON array of all {n_items} original 1-based item numbers in
recommended order.
Example format:
["CategoryA","CategoryB"]
[3,1,5,...]
```

**Figure S5.** AI Agent Choice Prompts – Batch Conditions

Generic

```
Below is a list of 150 {domain}. Which {n} {domain} would you recommend to
experience next?

Be sure to only select {n} {domain}, and respond with only the title of the
{item}. The title appears at the beginning of each entry before the colon. Don't
add any explanation. Your output should be structured as follows:
Title1, Title2, ..., Title{n}. Do NOT add any numbering, just the title.

Remember, you can ONLY select {domain} from this list below.

The 150 {domain} to choose from are:
{item_list}
```

Personalized (Summary)

```
Below is a summary description of a person:

{summary}

Please help this person select {n} {domain} from the list of 150 {domain} below.
Think carefully and consider what aligns most with their preferences. Which {n}
{domain} would you recommend they experience next?

Be sure to only select {n} {domain}, and respond with only the title of the
{item}. The title appears at the beginning of each entry before the colon. Don't
add any explanation. Your output should be structured as follows:

Title1, Title2, ..., Title{n}. Do NOT add any numbering, just the title.

Remember, you can ONLY select {domain} from this list below.

The 150 {domain} to choose from are:
{item_list}
```

Personalized (Domain)

```
Below are {domain} that a person has previously chosen:

{chosen_items}

Based on these preferences, please help this person select {n} NEW {domain} from
the list of 150 {domain} below. Do NOT select any {domain} the person has
already chosen. Think carefully and consider what aligns most with their
```

```
demonstrated preferences. Which {n} {domain} would you recommend they experience
next?

Be sure to only select {n} {domain}, and respond with only the title of the
{item}. The title appears at the beginning of each entry before the colon. Don't
add any explanation. Your output should be structured as follows:
Title1, Title2, ..., Title{n}. Do NOT add any numbering, just the title.

Remember, you can ONLY select {domain} from this list below, and you must NOT
repeat any of the already chosen {domain}.

The 150 {domain} to choose from are:
{item_list}
```

Personalized Hybrid (Domain + Summary)

```
Below is a summary description of a person:

{summary}

This person has also previously chosen the following {domain}:

{chosen_items}

Based on both their personality and their domain-specific preferences, please
help this person select {n} NEW {domain} from the list of 150 {domain} below. Do
NOT select any {domain} the person has already chosen. Think carefully and
consider what aligns most with their preferences. Which {n} {domain} would you
recommend they experience next?

Be sure to only select {n} {domain}, and respond with only the title of the
{item}. The title appears at the beginning of each entry before the colon. Don't
add any explanation. Your output should be structured as follows:
Title1, Title2, ..., Title{n}. Do NOT add any numbering, just the title.

Remember, you can ONLY select {domain} from this list below, and you must NOT
repeat any of the already chosen {domain}.

The 150 {domain} to choose from are:
{item_list}
```

**Figure S6.** AI Agent Choice Prompts – Incremental Conditions

Generic

```
Below is a list of {domain}. Which ONE {item} would you recommend next?

Respond with ONLY the title of the {item}, nothing else. The title appears at
the beginning of each entry before the colon.

Remember, you can ONLY select from this list below.

The {domain} to choose from are:
{item_list}
```

Personalized (Summary)

```
Below is a summary description of a person:

{summary}

Based on their personality, recommend ONE {item} from the list below that this
person would enjoy next.

Respond with ONLY the title of the {item}, nothing else. The title appears at
the beginning of each entry before the colon.

Remember, you can ONLY select from this list below.

The {domain} to choose from are:
{item_list}
```

Personalized (Domain)

```
Below are {domain} that a person has previously chosen:

{chosen_items}

Based on these preferences, recommend ONE new {item} from the list below that
this person would enjoy next.

Respond with ONLY the title of the {item}, nothing else. The title appears at
the beginning of each entry before the colon.

Remember, you can ONLY select from this list below.

The {domain} to choose from are:
{item_list}
```

Personalized Hybrid (Domain + Summary)

```
Below is a summary description of a person:

{summary}

This person has also previously chosen the following {domain}:

{chosen_items}

Based on both their personality and preferences, recommend ONE new {item} from
the list below that this person would enjoy next.

Respond with ONLY the title of the {item}, nothing else. The title appears at
the beginning of each entry before the colon.

Remember, you can ONLY select from this list below.

The {domain} to choose from are:
{item_list}
```